\documentclass[twocolumn,twocolappendix,Apj]{aastex63}
\usepackage[varg]{txfonts}
\usepackage{graphicx}
\usepackage[utf8]{inputenc}
\usepackage{hyperref}
\hypersetup{breaklinks,colorlinks,citecolor=blue,linkcolor=black,urlcolor=blue}
\usepackage{comment}
\usepackage{tabularborder}
\usepackage{xcolor}
\usepackage{natbib}
\defcitealias{nanni20}{N20}
\usepackage{amssymb}
\usepackage{comment}
\usepackage{enumitem}
\usepackage{gensymb}
\usepackage{wrapfig}

\newcommand\Tstrut{\rule{0pt}{2.7ex}}         % = `top' strut
\newcommand\Bstrut{\rule[-1.0ex]{0pt}{0pt}}   % = `bottom' strut

\shorttitle{X-ray redshifts for obscured AGN}
\shortauthors{Peca et al.}

\graphicspath{{./}{figures/}}

\begin{document}

\title{\Large{X-ray redshifts for obscured AGN: a case study in the J1030 deep field}}

%\correspondingauthor{Alessandro Peca}
%\email{alessandro.peca@miami.edu}

\author[0000-0003-2196-3298]{{Alessandro Peca}}
\affiliation{Department of Physics, University of Miami,  
Coral Gables, FL 33124, USA;
{\normalfont \href{mailto:alessandro.peca@miami.edu}{alessandro.peca@miami.edu}}}
\affiliation{INAF - Osservatorio di Astrofisica e Scienza dello Spazio di Bologna - 
via Gobetti 93/3, 40129 Bologna, Italy}
\affiliation{Dipartimento di Fisica e Astronomia, Universit\`a degli Studi di Bologna - 
via Gobetti 93/2, 40129 Bologna, Italy}

\author[0000-0002-8853-9611]{Cristian Vignali}
\affiliation{Dipartimento di Fisica e Astronomia, Universit\`a degli Studi di Bologna - 
via Gobetti 93/2, 40129 Bologna, Italy}
\affiliation{INAF - Osservatorio di Astrofisica e Scienza dello Spazio di Bologna - 
via Gobetti 93/3, 40129 Bologna, Italy}

\author[0000-0001-8121-6177]{Roberto Gilli}
\affiliation{INAF - Osservatorio di Astrofisica e Scienza dello Spazio di Bologna - 
via Gobetti 93/3, 40129 Bologna, Italy}

\author[0000-0002-9087-2835]{Marco Mignoli}
\affiliation{INAF - Osservatorio di Astrofisica e Scienza dello Spazio di Bologna - 
via Gobetti 93/3, 40129 Bologna, Italy}

\author[0000-0002-2579-4789]{Riccardo Nanni}
\affiliation{Department of Physics, University of California, Santa Barbara, CA 93106-9530, USA}
\affiliation{INAF - Osservatorio di Astrofisica e Scienza dello Spazio di Bologna - 
via Gobetti 93/3, 40129 Bologna, Italy}
\affiliation{Dipartimento di Fisica e Astronomia, Universit\`a degli Studi di Bologna - 
via Gobetti 93/2, 40129 Bologna, Italy}

\author[0000-0001-5544-0749]{Stefano Marchesi}
\affiliation{INAF - Osservatorio di Astrofisica e Scienza dello Spazio di Bologna - 
via Gobetti 93/3, 40129 Bologna, Italy}
\affiliation{Department of Physics and Astronomy, Clemson University - 
Kinard Lab of Physics, Clemson, SC 29634, USA}

\author{Micol Bolzonella}
\affiliation{INAF - Osservatorio di Astrofisica e Scienza dello Spazio di Bologna - 
via Gobetti 93/3, 40129 Bologna, Italy}

\author{Marcella Brusa}
\affiliation{Dipartimento di Fisica e Astronomia, Universit\`a degli Studi di Bologna - 
via Gobetti 93/2, 40129 Bologna, Italy}
\affiliation{INAF - Osservatorio di Astrofisica e Scienza dello Spazio di Bologna - 
via Gobetti 93/3, 40129 Bologna, Italy}

\author{Nico Cappelluti}
\affiliation{Department of Physics, University of Miami,  
Coral Gables, FL 33124, USA;
{\normalfont \href{mailto:alessandro.peca@miami.edu}{alessandro.peca@miami.edu}}}

\author[0000-0003-3451-9970]{Andrea Comastri}
\affiliation{INAF - Osservatorio di Astrofisica e Scienza dello Spazio di Bologna - 
via Gobetti 93/3, 40129 Bologna, Italy}

\author{Giorgio Lanzuisi}
\affiliation{INAF - Osservatorio di Astrofisica e Scienza dello Spazio di Bologna - 
via Gobetti 93/3, 40129 Bologna, Italy}

\author{Fabio Vito}
\affiliation{Scuola Normale Superiore, 
Piazza dei Cavalieri 9, Pisa, Italy}

\nocollaboration{12}

\begin{abstract}
\noindent We present a procedure to constrain the redshifts of obscured ($N_H > 10^{22}$ cm$^{-2}$) Active Galactic Nuclei (AGN) based on low-count statistics X-ray spectra, which can be adopted when photometric and/or spectroscopic redshifts are unavailable or difficult to obtain. 

\noindent We selected a sample of 54 obscured AGN candidates on the basis of their X-ray hardness ratio, $HR>-0.1$, in the \textit{Chandra} deep field ($\sim$479 ks, 335 arcmin$^2$) around the $z=6.3$ QSO SDSS J1030+0524. The sample has a median value of $\approx80$ net counts in the 0.5-7 keV energy band.
We estimate reliable X-ray redshift solutions taking advantage of the main features in obscured AGN spectra, like the Fe 6.4 keV K$\mathrm{\alpha}$ emission line, the 7.1 keV Fe absorption edge and the photoelectric absorption cut-off. The significance of such features is investigated through spectral simulations, and the derived X-ray redshift solutions are then compared with photometric redshifts.
Both photometric and X-ray redshifts are derived for 33 sources. When multiple solutions are derived by any method, we find that combining the redshift solutions of the two techniques improves the rms by a factor of two.
Using our redshift estimates ($0.1\lesssim z \lesssim 4$), we derived absorbing column densities in the range $\sim 10^{22}-10^{24}$ cm$^{-2}$ and absorption-corrected, 2-10 keV rest-frame luminosities between $\sim 10^{42}$ and $10^{45}$ erg s$^{-1}$, with median values of $N_H = 1.7 \times 10^{23}$ cm$^{-2}$ and $L_{\mathrm{2-10\, keV}} = 8.3\times10^{43}$ erg s$^{-1}$, respectively.
Our results suggest that the adopted procedure can be applied to current and future X-ray surveys, for sources detected only in the X-rays or that have uncertain photometric or single-line spectroscopic redshifts. 
\end{abstract}


\section{Introduction}

Active Galactic Nuclei (AGN) are the observed manifestation of gas accretion onto Super Massive Black Holes (SMBHs). The energy produced in this process can be observed from the radio frequencies to the X-rays and can dominate the host galaxy emission. However, the AGN radiation can be extinguished by gas and dust along our line of sight, making the detection of the AGN processes very challenging. 
Following the unified model for AGN (\citealt{antonucci93, urry95}), the presence of pc-scale, circumnuclear material distributed in a toroidal shape, may partially or completely hide the nuclear activity. In this case, the stellar emission from the host galaxy dilutes significantly the radiation produced by SMBH accretion, especially in the Optical/Near-InfraRed (ONIR) bands (e.g., \citealt{hickox18}), hiding the AGN from our view.
In this scenario, the high-energy X-ray photons can penetrate through high column densities making the detection of the obscured AGN ($N_H > 10^{22}$ cm$^{-2}$) possible. In addition, the X-ray radiation does not suffer from significant contamination, because of the very low contribution from stellar processes at typical AGN luminosity regimes ($L_X > 10^{42}$ erg s$^{-1}$; e.g., \citealt{padovani17}). X-ray surveys are therefore the best tool for revealing and characterizing the large population of obscured and faint AGN, which is predicted by X-ray background models (e.g., \citealt{comastri95,treister06,gilli07,ananna19}), but is the most challenging to detect.
In particular, to reveal mildly and heavily obscured ($N_H>10^{23}$ cm$^{-2}$) objects, very deep X-ray surveys (see \citealt{brandt15} for a review) are fundamental. 
It is worth mentioning that also the far-infrared/radio band is effective in selecting obscured AGN, but due to the relatively modest sensitivities of the current facilities, its potential is not yet fully exploited (e.g., \citealt{hickox18}). 

Obscured AGN are the most abundant class of objects revealed in deep X-ray surveys (e.g., \citealt{ueda03,aird15,buchner15}), and measuring their redshift is notoriously complicated, but at the same time crucial to understand their demography and their role in the AGN cosmological evolution.
ONIR spectroscopy is commonly used to provide the best redshift estimates, because of the uniquely identifiable emission and absorption features at these wavelengths. However, it is costly in terms of observing time and suffers from extinction, becoming not always feasible for faint sources. 
Photometry is then commonly used to build the Spectral Energy Distribution (SED) of such targets, providing a photometric redshift estimate ($z_{phot}$) whose accuracy depends on the data quality, the availability of suitable SED templates for the fitted objects, and the number of available filters.
Since for obscured AGN the radiation emitted by nuclear accretion is expected to be heavily suppressed in the ONIR bands, the photometric points are representative mostly of the stellar emission. Therefore, simple galaxy templates can be used for the SED fitting without the need of introducing hybrid (AGN + stellar) templates, which would produce degenerate redshift solutions (\citealt{salvato09}). 
However, especially for sources detected in the ONIR wavebands with a low signal-to-noise ratio and in a limited number of filters, photometric redshifts may be uncertain and not reliable (e.g., \citealt{salvato19}).

For these reasons, redshift estimates based on X-ray features in obscured AGN have been attempted (e.g., \citealt{maccacaro04,braito05,civano05,iwasawa12,vignali15}). This relatively new and promising technique relies on the main X-ray spectral features, like the Fe K$\alpha$ emission line and the Fe K$\alpha$ absorption edge, which become particularly prominent in heavily obscured objects (e.g., \citealt{ghisellini94,ikeda09}), allowing their identification and, consequently, the redshift estimate.
X-ray redshifts ($z_X$) were recently measured for hundreds of AGN in different surveys (e.g., \citealt{simmonds18,iwasawa20}) and also for galaxy clusters (e.g., based on the K–shell Fe line complex at 6.7-6.9 keV; \citealt{yu11}), with different approaches, selection criteria and photon statistics.
We explore here a method to constrain the redshifts of obscured AGN using low-count statistics X-ray spectra, down to $\sim30$ counts.
X-ray redshift solutions are derived from the combination of spectral analysis and ad-hoc spectral simulations, where the instrument response, a proper background sampling, and their off-axis dependencies are taken into account.
The obtained X-ray redshift solutions are also compared with photometric redshifts, derived from SED fitting. 
We show that the derived redshift quality is sufficient to calculate the main physical properties of obscured AGN, such as X-ray luminosity and absorption column density.
The proposed method can be applied to current X-ray surveys performed with {\it Chandra} and XMM-{\it Newton}, the forthcoming \textit{eROSITA} all-sky survey (eRASS) and to future X-ray missions.

The paper is organized as follows. In \S \ref{ch2} we present the multi-band data used in this work. In \S \ref{ch3} we describe the X-ray spectral analysis and the simulations used to estimate the X-ray redshifts. In \S \ref{ch4} we describe the SED fitting procedure. The main results are presented in \S \ref{ch5}, where we also show the derived AGN physical properties of our sample. Our results are discussed in \S \ref{ch6} and summarized in \S \ref{ch7}.
Throughout this paper, we assume a LCDM cosmology with the fiducial parameters $H_0 = 70$ km s$^{-1}$ Mpc$^{-1}$, $\Omega_M = 0.3$ and $\Omega_{\Lambda} = 0.7$, close to the Planck 2015 results (\citealt{planck16}). All magnitudes are in the AB system (\citealt{oke74}) and the errors are reported at the 1$\sigma$ confidence level if not specified otherwise.

\section{Dataset and sample selection} \label{ch2}
Our study uses data from the deep X-ray survey field centered on the $z=6.31$ quasar SDSS J1030+0524 (hereafter, the J1030 field; \citealt{nanni18}). This area was extensively covered by a large number of deep and wide multi-band observations (details on the J1030 web-page\footnote{{\url{http://j1030-field.oas.inaf.it/}}}).
A summary of the X-ray and optical/infrared datasets used in this work is reported below.

\begin{figure*}[!t]
	\begin{center}
		\includegraphics[width=0.97\textwidth]{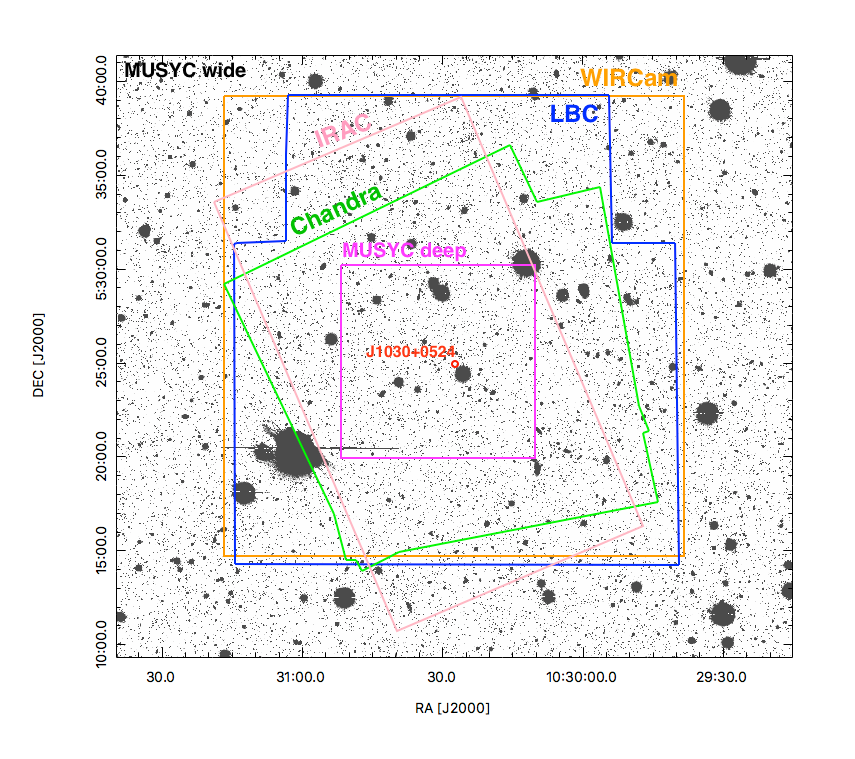}\hfil
		\caption{Multi-wavelength coverage of the J1030 field used in this work: \textit{Chandra}/ACIS-I (in green), LBT/LBC (in blue), CFHT/WIRCam (in orange), MUSYC K-deep (in magenta) and \textit{Spitzer}/IRAC from the IRSA online archive (in pink). The background image is the MUSYC $BVR$ stacked image, while the $z=6.31$ QSO SDSS J1030+0524 is highlighted in red.}
		\label{all_coverage}
	\end {center}
\end{figure*}

\subsection{Chandra observations}
The J1030 field was observed by \textit{Chandra}/ACIS-I with ten different pointings between January and May 2017, for a total exposure time of $\sim$479 ks and a field of view of 335 arcmin$^2$. This set of observations makes the J1030 field one of the deepest extragalactic X-ray survey performed so far, allowing us to investigate the obscured AGN population up to high-redshift and down to limiting fluxes of $\sim$3, 0.6, 2 $\times 10^{-16}$ erg s$^{-1}$ cm$^{-2}$ in the 0.5-7 keV (full), 0.5-2 keV (soft) and 2-7 keV (hard) bands, respectively.
Differently from the other \textit{Chandra} deep/moderately-deep fields (e.g., CDF-S, \citealt{giacconi01,luo17}; COSMOS-Legacy, \citealt{civano12,marchesi16b}), the J1030 field has not yet benefited from decades of spectroscopic follow-ups.
The details on the observations and data reduction are given in Section 2 of \citet{nanni18}. The source catalog (\citealt{nanni20}, hereafter \citetalias{nanni20}) has been generated using \textit{wavdetect} (\citealt{freeman02}) for the source detection and CIAO \textsc{Acis Extract} (\citealt{broos10}) for source photometry and significance assessment.
The final catalog contains 256 X-ray sources. These have then been matched with the available optical/infrared catalogs using a likelihood ratio technique (e.g., \citealt{ciliegi03, brusa07, luo10}); 252 of them have a counterpart in these wavebands (see \citetalias{nanni20} for further details).

\subsection{Optical/infrared imaging}
In 2012 the J1030 field was observed with the Large Binocular Telescope using the Large Binocular Camera (LBT/LBC)  to obtain imaging of a $23\arcmin \times 25\arcmin$ area in the $r$, $i$ and $z$ bands (\citealt{morselli14}), down to limiting AB magnitudes of 27.5, 25.5, and 25.2, respectively.
In 2015 we performed a $24\arcmin \times 24\arcmin$ observation in the near-infrared $Y$ and $J$ bands (\citealt{balmaverde17}) at the Canada France Hawaii Telescope using WIRCam (CFHT/WIRCam), with limiting AB magnitudes of 23.8 and 23.75, respectively.

The field is one of the four fields included in the Multi-wavelength Survey by Yale-Chile (MUSYC). Three MUSYC catalogs are available for J1030: the BVR catalog, obtained by selecting sources in the $BVR$ stacked image down to 26.3 AB magnitudes (\citealt{gawiser06}),  the K-wide (\citealt{blanc08}) and K-deep (\citealt{quadri07}) catalogs, performed selecting sources in the $K$ band down to $K$ = 21 and 23 AB, respectively. As shown in Figure \ref{all_coverage}, the MUSYC BVR and K-wide data cover a $30\arcmin \times 30\arcmin$ area, while the MUSYC K-deep covers a smaller $10\arcmin \times 10\arcmin$ area.
Furthermore, the field has been observed by the \textit{Spitzer} Infrared Array Camera (IRAC) in the Mid-InfraRed (MIR) at 3.6 and 4.5 $\mu m$. In this work we used the available catalogs and images in the IRSA\footnote{{\url{https://irsa.ipac.caltech.edu/}}} archive, that reach a depth of 22-23 AB magnitudes. 
All the optical/infrared datasets and filters used in this paper are discussed in Section \ref{ch4}.

\subsection{Sample selection} \label{sample_selection}
\begin{figure}[t!]
	\begin{center}
		\includegraphics[width=0.51\textwidth]{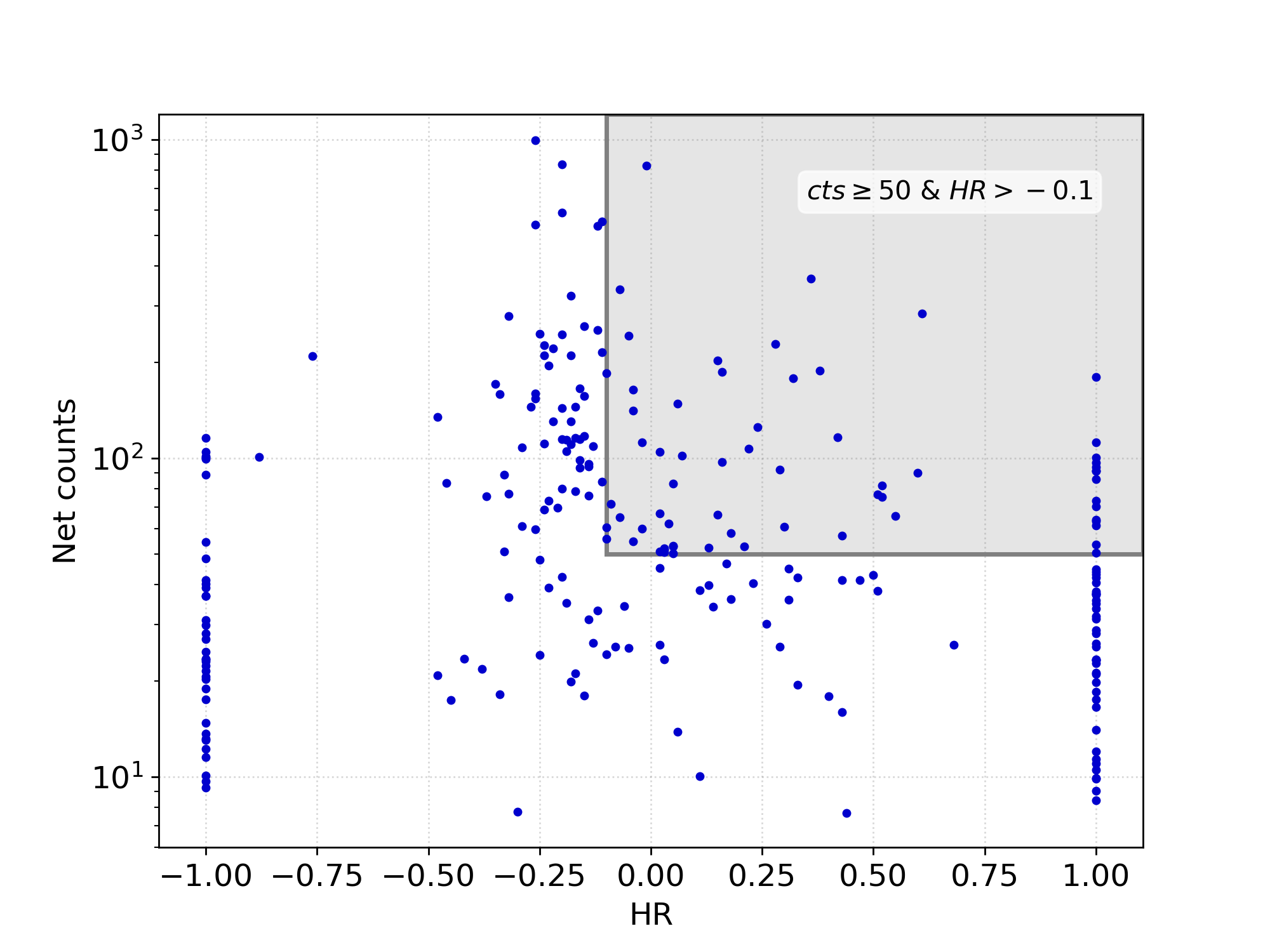}\hfil
		\caption{$HR$ versus full band net counts from the J1030 X-ray catalog. The grey top right corner shows the region with $HR>-0.1$ and net counts $\geq50$ used to select our sample of 54 obscured AGN candidates.}
		\label{HR_cts}
	\end {center}
\end{figure}

We selected, from the X-ray catalog, a sample of 54 obscured AGN candidates on the basis of their Hardness Ratio ($HR$), defined as
\begin{equation}\label{hr_eq}
	HR = \frac{H-S}{H+S}	
\end{equation}
where $H$ and $S$ are the net count rates (i.e. background subtracted) in the hard and soft bands, respectively.
Since the AGN X-ray emission is more absorbed at low energies, the $HR$ value can be considered a good proxy of absorption for AGN with known redshift and simple (absorbed power-law) spectra (e.g., \citealt{mainieri02}).
\begin{figure}[t!]
	\begin{center}
		\includegraphics[width=0.48\textwidth]{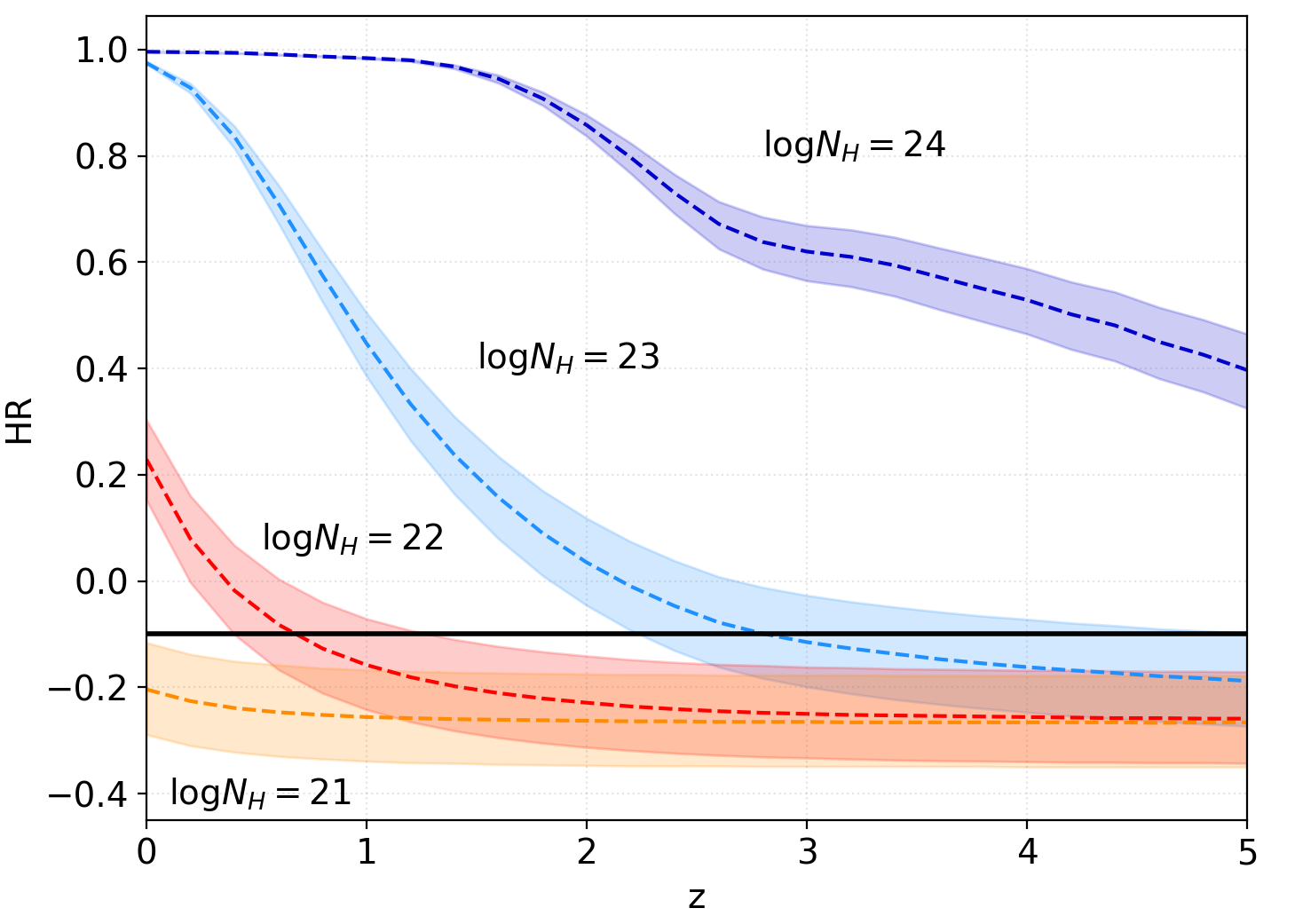}\hfil
		\caption{Hardness Ratio ($HR$) as a function of redshift, for different absorption column densities in color code. The shaded areas indicate $HR$ values derived with fixed $\Gamma=1.7$ (top curves) and $\Gamma=2.1$ (bottom curves), while the dotted lines represent models with $\Gamma=1.9$. We used the response matrices at the aimpoint of the \textit{Chandra} observations.
		The black horizontal line represents the chosen selection threshold $HR=-0.1$.}
		\label{tozzi_HR}
	\end {center}
\end{figure}
However, the \textit{Chandra}/ACIS-I photon collecting efficiency is rapidly decreasing  in the soft band due to contamination  (see the \textit{Chandra} Proposers' Observatory Guide\footnote{ \url{https://cxc.harvard.edu/proposer/POG/}}), making the hardness ratio a time-dependent quantity. This decrease has   accelerated over the last few years, thus allowing only a qualitative comparison with previous works (e.g., \citealt{tozzi01}; \citealt{szokoly04}) as explained in Appendix \ref{app_A}. Therefore, we performed X-ray spectral simulations based on the J1030 \textit{Chandra} observations to reproduce the expected trends for our sample. The simulations were performed through XSPEC\footnote{{\url{https://heasarc.gsfc.nasa.gov/xanadu/xspec/}}} v.12.9.1 (\citealt{arnaud96}), assuming an absorbed power-law model at different redshifts (\texttt{zphabs$\times$powerlaw}). The mean Galactic absorption at the J1030 field position, $N_H = 2.6\times 10^{20}$ cm$^{-2}$ (\citealp{kalberla05}), was also considered (\texttt{phabs}).
In Figure \ref{tozzi_HR} we show the $HR$ values for typical AGN column densities ($10^{21} \leq N_H \leq 10^{24}$ cm$^{-2}$) as a function of redshift and with a canonical photon index $\Gamma=1.9\pm0.2$ (e.g., \citealt{nandra94, piconcelli05, lanzuisi13}).
It is worth mentioning that these curves are computed for simulated spectra with thousands of counts to reproduce the expected $HR$ trends (e.g., \citealt{szokoly04, elvis12}).
Based on our simulations, we chose a threshold of $HR > -0.1$ (black horizontal line) to select obscured AGN. In fact, considering a $\Gamma=1.9$, this threshold allows the selection of obscured objects with $N_H \geq 10^{22}$ cm$^{-2}$ up to $z\approx 0.5$, $N_H \geq 10^{23}$ cm$^{-2}$ up to $z\approx 2.5$ and Compton-thick AGN ($N_H \gtrsim 10^{24}$ cm$^{-2}$) at all redshifts. If instead, we consider a flatter ($\Gamma = 1.7$) or a steeper ($\Gamma = 2.1$) power-law, we obtain more positive or more negative $HR$, respectively, but, in both cases, the chosen $HR> -0.1$ threshold avoids the selection of unobscured sources at any redshift.
Due to the limited photon statistics in our sample, we assumed a basic model to compute the different $N_H$ curves shown in Figure \ref{tozzi_HR}.  Despite this, if we consider more complex models, the chosen $HR$ threshold remains valid (see Appendix \ref{app_B}).
In addition to the $HR$ criterion, we selected sources detected in \citetalias{nanni20} with at least 50 net counts in the 0.5-7 keV band to allow an effective search for X-ray spectral features, such as the 6.4 keV Fe K$\alpha$ emission line and the 7.1 keV Fe absorption edge. 
For sources not detected in one or two bands, \citetalias{nanni20} reported the 3$\sigma$ net counts upper limits. Because we were looking for relatively hard objects, we considered a $HR=1$ for sources undetected in the soft band, and $HR=-1$ for sources undetected in the hard band. We discarded 11 sources detected in the full band only.
The final sample and the adopted selection criteria are shown in Figure \ref{HR_cts}.

\section{X-ray data analysis} \label{ch3}
\subsection{Spectral analysis} \label{xray_spec}
The spectral extraction was made using the Chandra Interactive Analysis of Observations\footnote{{\url{http://cxc.harvard.edu/ciao/}}} (CIAO) v.4.9 software.
The choice of the extraction regions was performed by taking into account the sources position on the detector, since the PSF broadens as the off-axis angle ($\theta$) increases. 
As extraction radius we use the 90\% encircled energy radius ($E=1.49$ keV) at the source position and, to increase the signal-to-noise ratio of the faintest sources, we manually chose ad-hoc slightly smaller radii.
The regions used to extract the background spectra were selected next to each source and, to ensure a good background sampling, with an area at least 10 times larger.
We extracted a spectrum from each observation covering a given source, and then combined these spectra using the CIAO tool \textsc{combine\_spectra}.
The source spectra were grouped to a minimum of one count per energy bin to avoid empty channels. We checked the presence of at least one background count in each source energy bin, and then adopted the modified C-statistic for direct background subtraction (\citealt{cash79,wachter79}), or W-statistic, to estimate the best-fit model parameters. The spectral analysis was performed using XSPEC v.12.9.1 (\citealt{arnaud96}).

\begin{figure}[t!]
    \begin{center}
        \includegraphics[width=0.5\textwidth]{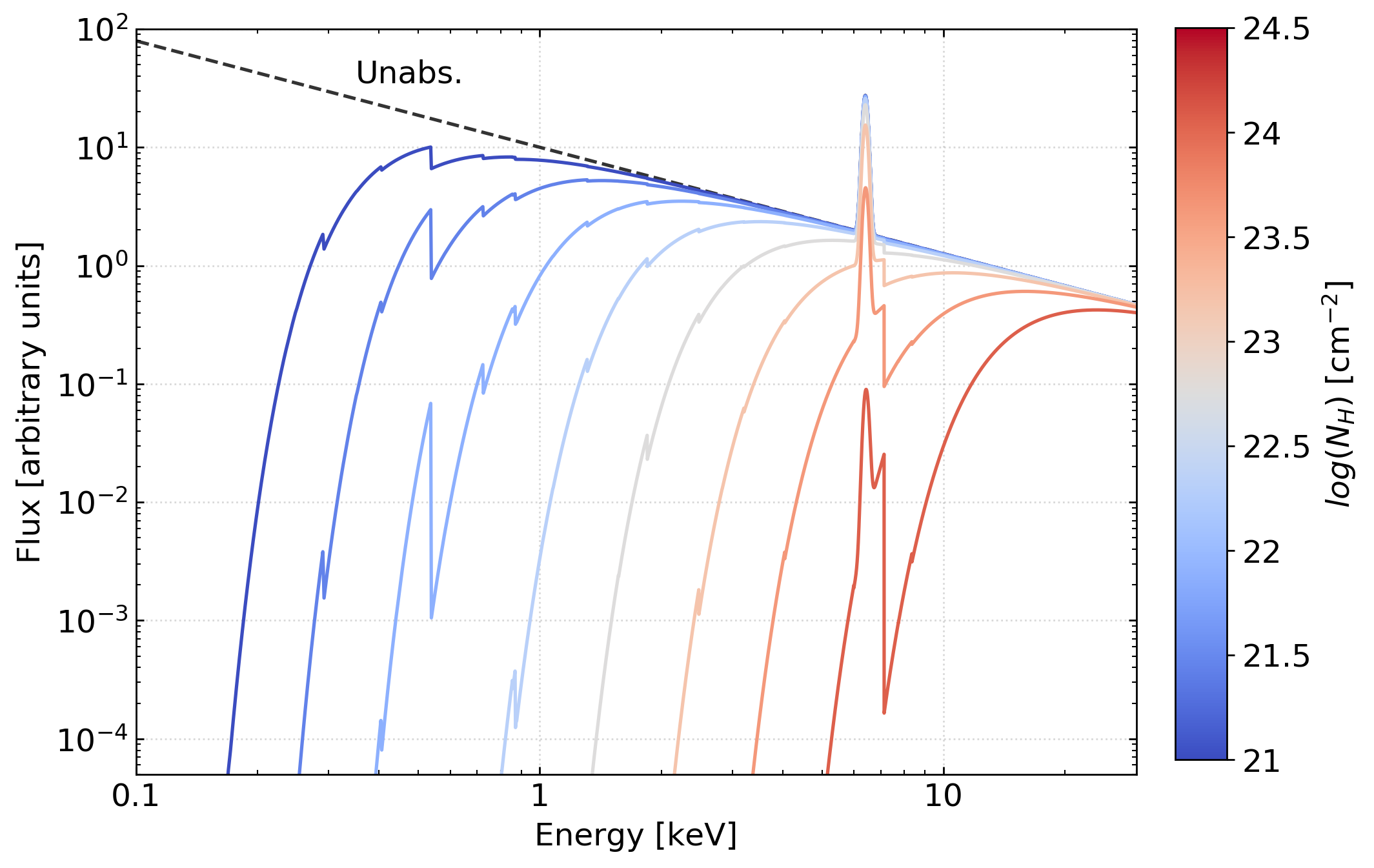}\hfil
        \caption{X-ray AGN spectral model as a function of the absorption column density, $N_H$, in color code. The assumed model is a power-law (\texttt{powerlaw}) with $\Gamma=1.9$, a fixed Gaussian line at 6.4 keV rest-frame with $\sigma=10$ eV (\texttt{zgauss}), and an absorption component (\texttt{zphabs}). The plotted curves are for $z=0$.}
        \label{simil_gilli07}
    \end {center}
\end{figure}

\begin{figure}[t!]
	\begin{center}
		\includegraphics[width=0.5\textwidth]{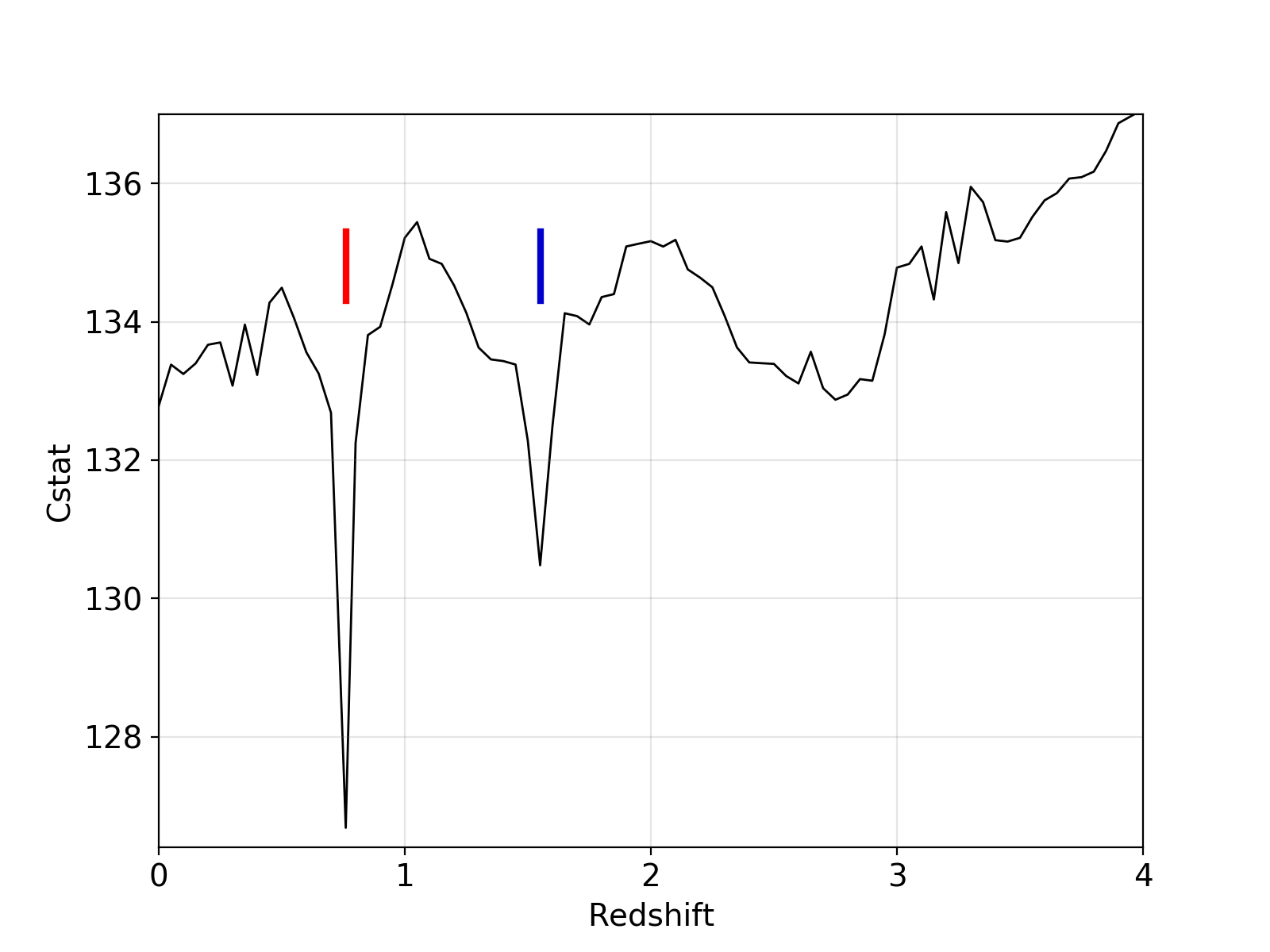}\hfil
		\caption{Redshift likelihood profile in terms of C-statistic, for one source (XID 70) of the selected sample. In red and blue are shown the primary and secondary $z_X$ solutions, respectively. The profile was computed through the \textsc{steppar} command with a redshift step of $\Delta z=0.05$.}
		\label{cstatdistrib}
	\end {center}
\end{figure}

Our sample of 54 objects has a median value of $\approx$80 net counts in the 0.5-7 keV energy range, and median fluxes of $6.8,1.0,5.4\times 10^{-15}$ erg s$^{-1}$ cm$^{-2}$ in the full, soft and hard bands, respectively.
Given the low photon statistics, we adopted a simple model based on a power-law, an intrinsic $N_H$ at the source redshift $z$, and Galactic absorption at the source position (\texttt{phabs(zphabs$\times$powerlaw)}). The photon index $\Gamma$ was fixed to 1.9 as commonly observed in AGN (e.g., \citealt{nandra94, lanzuisi13}). We let the intrinsic $N_H$, $z$, and the power-law normalization free to vary.
To investigate the presence of emission lines, we included a redshifted Gaussian feature at 6.4 keV rest-frame (\texttt{zgauss}) with a redshift parameter anchored to the corresponding absorption component, a free normalization, and a fixed line width $\sigma = 10$ eV (e.g., \citealt{nanni18}), which takes into account only the narrow component produced far away from the central SMBH by the absorbing cold medium, since the broad, relativistic component produced in the accretion disk is expected to be obscured (e.g., \citealt{risaliti04}).
Using a simple absorbed power-law plus a Gaussian line may not be a detailed description of the X-ray spectrum. Nevertheless, the limited count statistics did not allow an investigation of more complex spectral shapes (e.g., \citealt{lanzuisi13,iwasawa20}). This choice is also justified by a few tests on more complex models, showed in Appendix \ref{app_B2}.
In case of heavy obscuration the main AGN features, such as the Fe K$\alpha$ 6.4 keV line and the 7.1 keV Fe edge, become more prominent (e.g., \citealt{iwasawa12}) and more easily recognizable, as shown in Figure \ref{simil_gilli07}.
Once one of these features was identified, a redshift solution from the X-ray spectrum (hereafter, $z_X$) was derived by evaluating the redshift likelihood profile, computed with the \textsc{steppar} command. An example is reported in Figure \ref{cstatdistrib}. We considered reliable $z_X$ those solutions where the difference in C-statistic ($\Delta C$) between the global minimum (primary solution, i.e. the best-fit redshift) and its nearby maximum is at least 2.71, corresponding to a fit improvement at the 90\% confidence (see e.g., \citealt{tozzi06, brightman14} who validated this threshold through simulations).
We also investigated local minima (secondary solutions) where the above $\Delta C$ criterion was satisfied, as for the case in Figure \ref{cstatdistrib}.
Each selected $z_X$ was then further investigated through simulations.

\subsection{X-ray spectral simulations} \label{xray_sim}
Estimating the goodness of an X-ray spectral fit is not trivial in case of low-count statistics. We therefore built two different sets of spectral simulations to test the derived X-ray redshift solutions.
The first set of simulations aims at verifying the significance of the candidate emission lines (Section \ref{lin_sim}) and the second at constraining for which photon statistics, as a function of $N_H$, $z$ and $\theta$, we expect to obtain robust redshifts from X-ray data alone (Section \ref{semitheor_sim}).
All simulated spectra were obtained using the XSPEC \textsc{fakeit} command.

\subsubsection{Line significance} \label{lin_sim}
The simulations reported in this paragraph refer to sources where the redshift estimate is driven by the iron K$\alpha$ emission line.
For each source in which the line was possibly detected, we established a significance criterion to deem a candidate emission line as reliable as follows (e.g., \citealt{lanzuisi13_2,vignali15}).
Because we do not know the redshift of the sources, we fitted the spectra using an absorbed power-law model (\texttt{phabs $\times$ powerlaw}, $\Gamma=1.9$) with and without a Gaussian line (\texttt{gauss}, $\sigma=10$ eV). For the former, the line energy was fixed to the observed value.
The best-fit parameters from the model without the line were then used to simulate 1000 spectra with the same characteristics (response matrices\footnote{Ancillary Response File (ARF) and Redistribution Matrix File (RMF).}, exposure time, background and photon statistics) of the observed one. 
We fitted the simulated spectra using exactly the two models, fixing the continuum to the one derived from the model without the line and leaving the line energy free to vary,
looking for all cases where 
\begin{equation}\label{deltac_eq}
	\Delta C_{sim} \geq \Delta C_{obs},
\end{equation}
where $\Delta C_{obs}$ is the difference between the best-fit model with and without the possible line in the observed spectrum, while $\Delta C_{sim}$ is the difference between the same models in each simulated spectrum.
The frequency at which (\ref{deltac_eq}) occurs corresponds to the probability $P'$ that the detected line is just a statistical fluctuation. Then 
\begin{equation}
P_{sim}=(1-P')
\end{equation}
corresponds to the significance of the observed line. We considered an emission line reliable when $P_{sim} \geq 90\%$.

Since calculating $P_{sim}$ is time-consuming, we proceeded as follows.
The F-test is a fast method to evaluate the model improvement due to an additive component, but in case of a Gaussian line it is considered inappropriate (\citealt{protassov02}). However, allowing the line normalization to be also negative\footnote{ {\url{https://asd.gsfc.nasa.gov/XSPECwiki/statistical_methods_in_XSPEC}}}, it can be used as an indication of the model improvement.
After an extensive testing on our dataset, we found that, to obtain a reliable line (i.e., with $P_{sim}\geq90\%$), an F-test probability ($P_{Ft}$) $>99\%$ is required. We then decided to use the F-test as pre-screening. When the $P_{Ft}$ threshold is reached, the significance ($P_{sim}$) of the candidate lines is computed through the aforementioned simulations, to provide a more solid evaluation.
When a significant line has been found, it is used to derive an X-ray redshift solution, assuming that the detected line is the Fe K$\alpha$ fluorescent emission line at 6.4 keV, which is the most probable emission line in the AGN X-ray spectrum (e.g., \citealt{fabian00}). Otherwise, the  $z_X$ determination is principally driven by the Fe 7.1 keV edge coupled with the photoelectric absorption cut-off.

\vspace{1mm}
\subsubsection{Redshift solutions as a function of $N_H$, net counts and off-axis angle} \label{semitheor_sim}
To verify the level of photon statistics allowing the redshifts to be derived from the X-ray analysis, we performed a second set of simulations. We simulated spectra not only with different AGN parameters, but also using different responses and backgrounds, as these vary with the off-axis angle. This set of simulations aims to be global, i.e. reliable in every position of the observations, and it was used to further investigate the derived X-ray redshift solutions.
To test $z_X$ driven by absorption features, we adopted an absorbed power-law model (\texttt{zphabs $\times$ powerlaw}) with a fixed intrinsic photon index $\Gamma = 1.9$, while for redshift solutions driven by the Fe K$\alpha$ line we also included a redshifted Gaussian line (\texttt{zgauss}) at 6.4 keV rest-frame, with a width of $\sigma = 10$ eV. We set different line normalizations to obtain a canonical range of rest-frame equivalent widths, between 10 eV and 2 keV, as a function of $N_H$ (e.g., \citealt{ghisellini94,lanzuisi15}). In both models we added an additional absorption component (\texttt{phabs}) with a fixed value of $N_H = 2.6 \times 10^{20}$ cm$^{-2}$, corresponding to the mean Galactic absorption at the J1030 field position.
To reproduce what is observed in deep X-ray surveys (e.g., \citealt{tozzi06, marchesi16b}), we simulated column densities from $N_H=10^{21}$ to $N_H=10^{24}$ cm$^{-2}$ with a logarithmic step of 0.5, redshifts up to 5 with a step of 0.5, and different power-law normalizations to obtain a number of full band net counts in the range 10-1000. For each parameter combination 500 spectra were simulated.

\begin{figure}[t!]
    \begin{center}
        \includegraphics[width=0.5\textwidth]{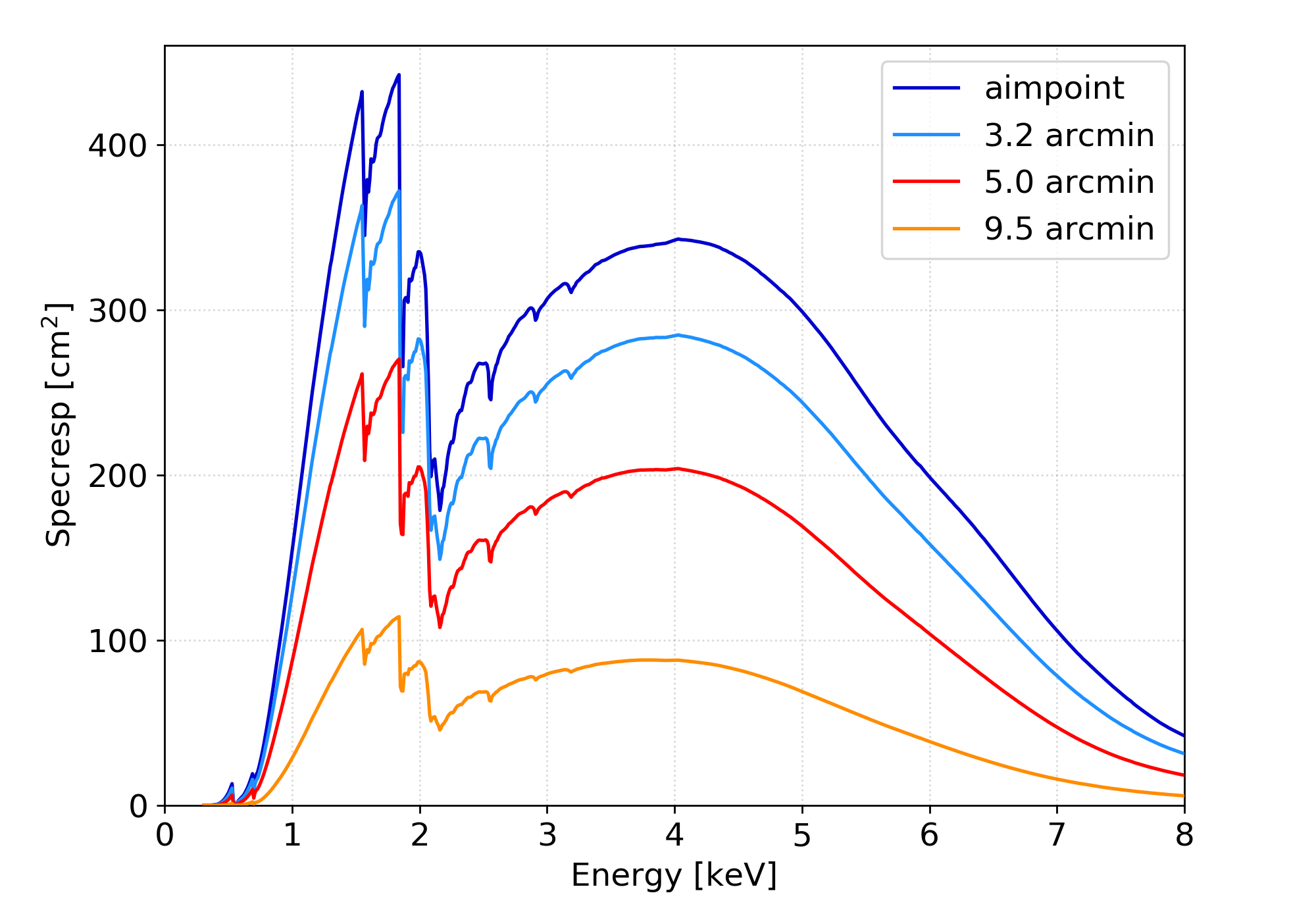}\hfil
        \caption{ARFs selected for the simulations at different off-axis angles from the aimpoint. As discussed in the text, the instrument's response decreases as the off-axis angle increases.}
        \label{arfs_offaxis}
    \end {center}
\end{figure}

To simulate spectra that are as close as possible to those observed in the \textit{Chandra} data, the response matrices of the real observation must be used.
In general, the instrumental response drops as the off-axis angle increases, which has therefore to be taken into account. Assuming no azimuthal dependence of the instrument response, we extracted the ARF (and RMF) for four relatively bright sources at different off-axis angles ($\theta \sim$0, 3.2, 5.0 and 9.5 arcmin) in our sample, in order to reproduce the decrease of the effective area as a function of $\theta$ (Figure \ref{arfs_offaxis}).
In addition, it is necessary to evaluate background spectra to be associated with the simulated source spectra. Assuming that the background only varies with the off-axis angle and has no azimuthal dependence, we extracted background spectra using a circular region of $r=2\arcmin$ centered at the aimpoint, and annuli of width 1$\arcmin$ at a distance of 3.5$\arcmin$, 5.5$\arcmin$, and 9.5$\arcmin$ from the aimpoint. To get a suitable background, the sources present in such regions were excluded. 
We used the CIAO tool \textsc{dmcopy}, through the command \textsc{exclude}, to remove  X-ray sources identified by \citetalias{nanni20} in the J1030 field.
The extracted background needs also to be rescaled by the source extraction area before being associated with it. Since the sources are simulated, they do not have a physical extraction region, so we rescaled the background to the width of the PSF (90\% encircled energy radius at $E=1.49$ keV) at specific off-axis angles.

\begin{figure}[t!]
    \begin{center}
        \includegraphics[width=0.49\textwidth]{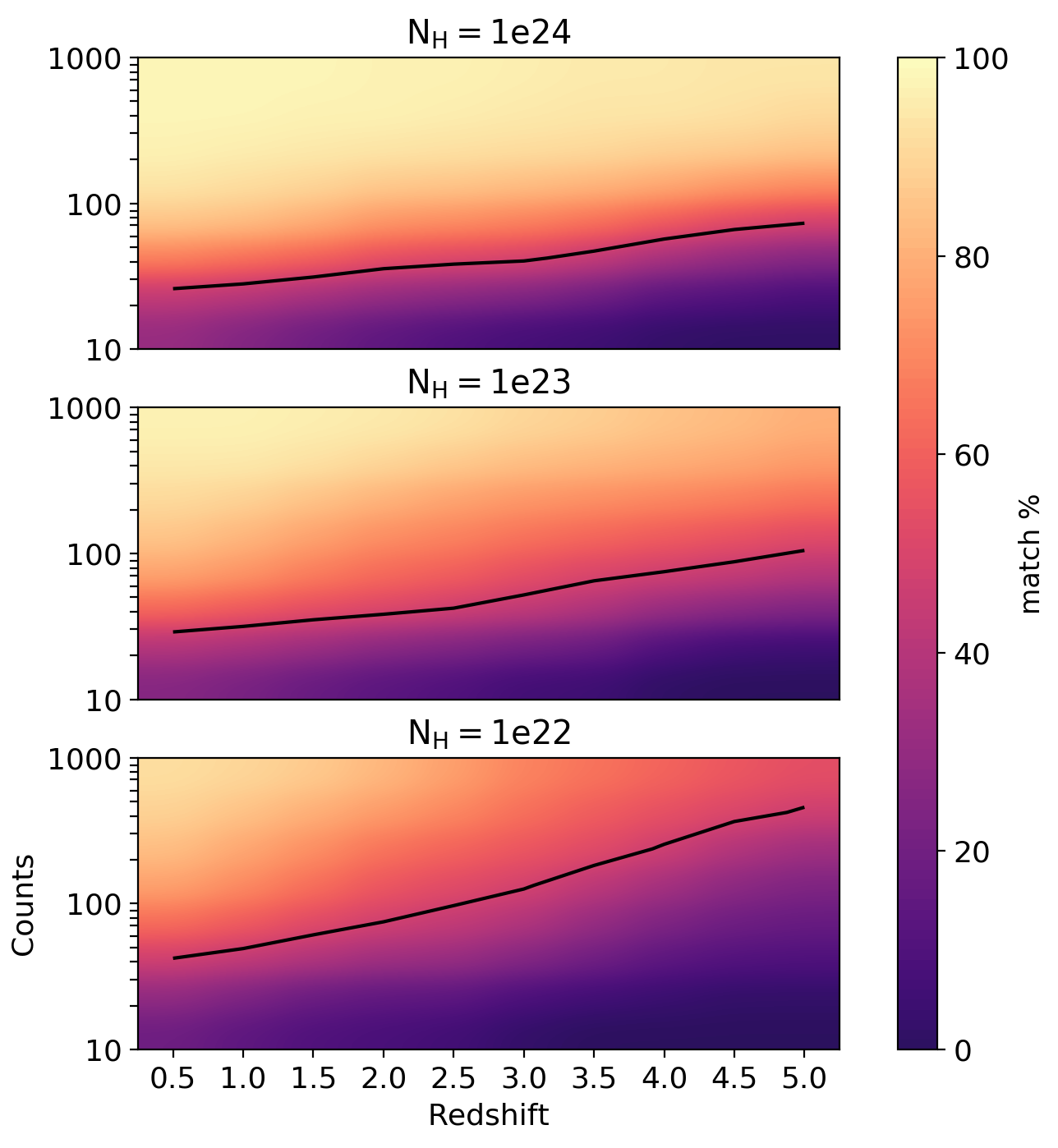}\hfil
        \caption{Simulation results for a source located within 2$\arcmin$ from the aimpoint, where the redshift solutions are derived using only absorption features. The three panels indicate different absorption values: $N_H = 10^{24}, 10^{23}$, and $10^{22}$ cm$^{-2}$, from top to bottom, respectively. Each panel shows the match percentage (Equation \ref{matcheq}) in color code, smoothed for graphical purposes, as a function of redshift and net counts. The black solid lines represent the 50\% confidence curves above which $z_X$ are considered reliable solutions. The simulation has a resolution of 20 net counts.}
        \label{sim_prova}
    \end{center}
\end{figure}

\begin{figure}[t!]
    \begin{center}
        \includegraphics[width=0.45\textwidth]{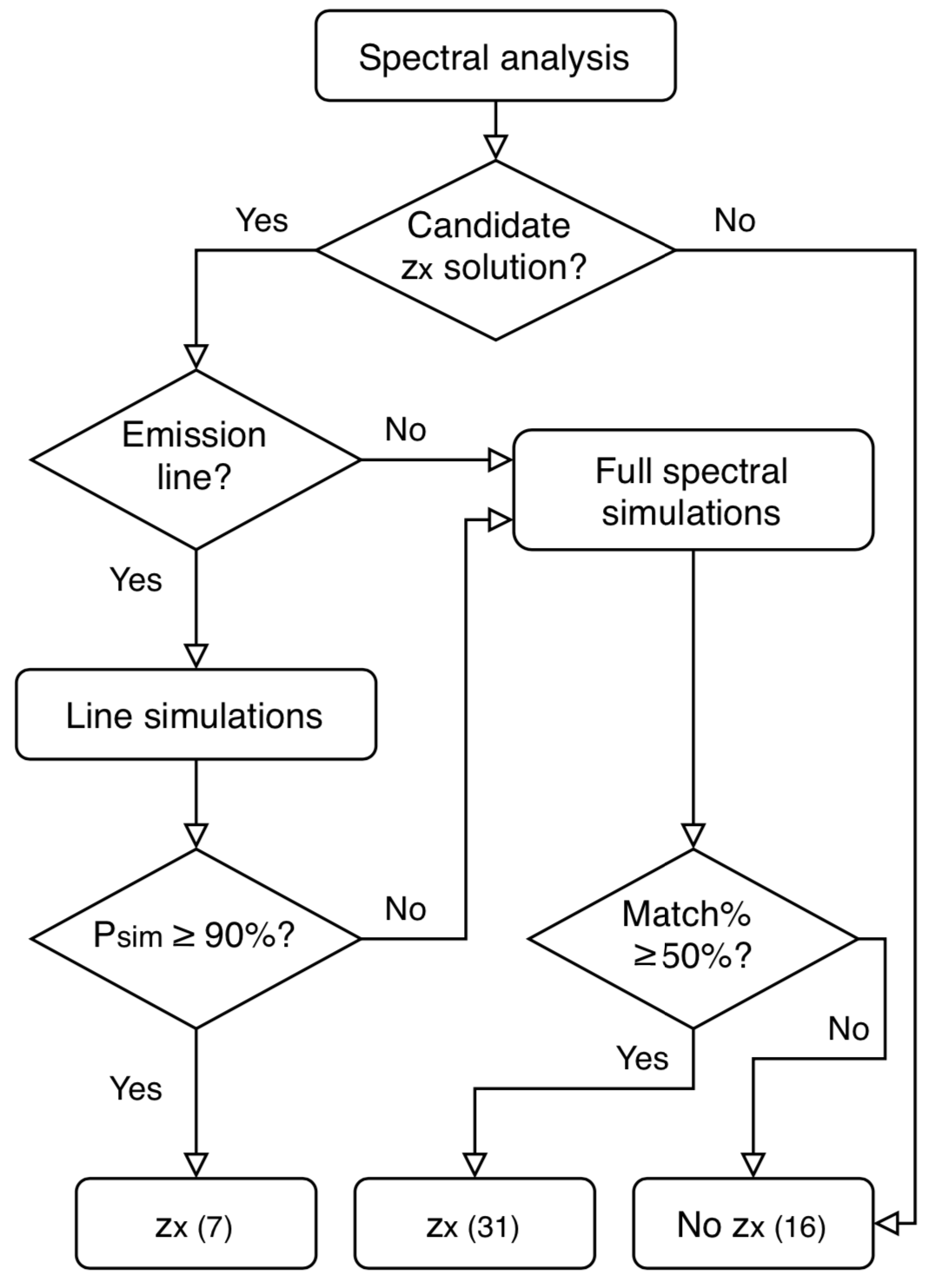}\hfil
        \caption{Flow chart of the adopted X-ray procedure. When an X-ray redshift solution was found through the spectral analysis, we firstly evaluated the significance of possible emission lines (Line simulation). If $P_{sim} \geq 90\%$, then we got a $z_X$ solution from the emission line (7 sources), otherwise the redshift solution is driven by absorption features. In this case we checked the redshift reliability through the match\% (Equation \ref{matcheq}, Full spectral simulations) with a threshold of $\geq 50\%$, that gave 31 reliable $z_X$ solutions. We did not find an X-ray redshift solution for 16 sources.}
        \label{xflow}
    \end{center}
\end{figure}

For each parameter combination ($N_H$, $z$, net counts and  $\theta$) 1000 spectra were simulated, for a total of 800,000. For each simulated spectra a fit was performed and the best-fit redshift solution, $z_X$, derived. We then computed the match percentage:
\begin{equation}\label{matcheq}
    \mathrm{match \%}\ (z,N_H,cts,\theta) = \frac{N(z_X \pm \Delta z)}{N(z_{sim})} 	
\end{equation}
that corresponds, for a specific range of redshift, $N_H$, number of net counts and $\theta$, to the number of simulated spectra in which $z_X$ is consistent with the simulated redshift ($z_{sim}$), within a given tolerance $\Delta z$, normalized to the total number of simulated spectra.
Because of the low-count statistics, the X-ray redshift solutions are sometimes poorly constrained. To determine if the redshift solutions are reliable, we rejected solutions with $|\Delta z| > 0.15(1+z_{sim})$, defined as outliers (this value has been found in previous works to be a reliable boundary for outliers, e.g., \citealt{hsu14, ananna17, luo17, simmonds18}). For a conservative approach, we discarded redshift solutions that are either upper or lower limits. We also checked the $N_H$ values, rejecting solutions that are not consistent within the errors with the simulations, even if the redshift solutions are good ($\sim 6\%$).
An example of the simulations' performance is shown in Figure \ref{sim_prova}, where it is clear how column densities $\geq 10^{22}$ cm$^{-2}$ are needed to obtain reliable $z_X$ for sources with a net counts range as in our sample. Furthermore, the match percentage depends on redshift. If no emission lines are detected, the X-ray solutions are driven by the iron absorption edge coupled with the photoelectric cut-off but, moving towards high-redshift ($z>$3-4), such absorption complex ends up at $\lesssim1.5$ keV, where the effective area decreases dramatically. As a consequence, the match percentage decreases as the redshift increases.
Besides showing how the main features of the X-ray spectrum become more prominent with increasing obscuration, and therefore more easily identifiable, the simulations give us an indication of the probability of deriving a correct redshift or not. 
We set a match percentage threshold of at least $50\%$ to accept $z_X$ solutions driven by absorption features. In this regard, we found reliable solutions down to a regime of $\sim$30 net counts, for redshift $\lesssim 2.5$ and $N_H > 10^{23}$ cm$^{-2}$. 
For $z_X$ solutions derived from the Fe K$\alpha$ line, instead, we considered this $50\%$ threshold as an additional check, since it confirms the results obtained from Section \ref{lin_sim}.
On our sample, where it was possible to derive an X-ray redshift solution (38 sources), the chosen threshold returns a mean match percentage of $\sim 70\%$.
The full procedure adopted for the $z_X$ estimate is summarized in Figure \ref{xflow}.

\vspace{9mm}
\section{Photometric data analysis} \label{ch4}

\subsection{Data modelling} \label{data_hz}
We used the available datasets in the optical and infrared bands to calculate photometric redshifts (hereafter, $z_{phot}$) through a SED fitting procedure, and test the X-ray redshift solutions.
Photometric redshifts were obtained through the \textit{hyperz} code (\citealt{bolzonella00}) using a variety of galaxy templates, detailed below, and a \citet{calzetti00} reddening law.
The code finds the best-fit template through a standard $\chi^2$ minimization procedure, comparing template spectra to the observed SEDs as:
\begin{equation}
\chi^2(z) = \sum^{N_{filters}}_{i=1} \Bigg[  \frac{F_{obs,i} - b\times F_{temp,i}(z)}{\sigma_i}  \Bigg]^2	
\end{equation}
where $F_{obs,i}$ and $F_{temp,i}$ are the observed and template fluxes, $\sigma_i$ is the observed flux uncertainty and $b$ is a normalization constant. \textit{hyperz} provides primary and secondary $z_{phot}$ solutions, the best-fitting template spectrum, and the reduced $\chi^2$ for each given object.
For a detailed description of the code we refer to \citet{bolzonella00}.
An essential requirement for the SED-fitting procedure is an extensive template library which covers the entire range of selected objects, without adding unphysical
degeneracies. 
Since the strong ONIR radiation from the accretion disk is heavily extinguished in obscured AGN, allowing the stellar emission of the host galaxy to dominate at these wavelengths (e.g., \citealt{merloni14}), photometric redshifts can be reliably estimated using standard galaxy templates.
In this regard, we included the available photometry up to $\sim5$ $\mu m$. Above this wavelength, in fact, we expected that the rest-frame emission of the hot dusty torus may overcome the host galaxy stellar emission (e.g., \citealt{pozzi12,circosta19}) invalidating the choice of galaxy templates.

\begin{table}[!t]
	\centering\small
	\begin{tabular}{cccc} \toprule
		Catalog           & Filters                &  Area            & Depth {[}AB{]} \\ \midrule
		MUSYC BVR         & $UBVRIz$          &  30'$\times$30' & 25-26                 \\
		LBT/LBC           & $riz$          &  23'$\times$25' & 27.5, 25.5, 25.2         \\
		CFHT/WIRCam       & $YJ$               &  24'$\times$24' & 23.8, 23.75              \\
		MUSYC K wide      & $UBVRIzK$        &  30'$\times$30' & 21                    \\
		MUSYC K deep      & $UBVRIzJHK$    &  10'$\times$10' & 23                    \\
		\textit{Spitzer}/IRAC      & ch1 ch2               &  35'$\times$35' & 22-23                 \\
		\bottomrule
	\end{tabular}
	\caption{Photometric catalogs used in our analysis. From the left: catalog name, filters, covered area and approximated limiting AB magnitudes.}
	\label{photo_data}
\end{table}

Following \citet{ilbert13}, we used a library composed by 75 galaxy templates: 19 empirical templates derived from the SWIRE library (\citealt{polletta07}), including both elliptical and spiral galaxies (S0, Sa, Sb, Sc, Sd, Sdm), plus 12 starburst templates and 44 additional red galaxy templates generated by \citet{ilbert09, ilbert13} through the \citet{bruzual03} stellar synthesis models, to better cover the color-redshift space.
We built source SEDs using the catalogs described in Table \ref{photo_data}: a total of 12 different filters are available from the blue-optical to the MIR wavelengths. When the same source is revealed in the same filter in more than one catalog, we used the magnitude value from the deepest observation, while if the source is either not detected in a specific band or it is detected with a signal-to-noise ratio $<2$, we excluded the corresponding filter from the SED fitting procedure.
We searched photometric redshift solutions from $z=0$ to $z=7$, with a step $\Delta z = 0.05$.
The photometric redshift accuracy depends also on the number of filters in which a specific source is detected. %We set a threshold of at least five filters, which gives an rms$\approx$0.1.
In this regard, Marchesi et al. (in prep.) carried out a photometric redshift analysis for all the X-ray sources in the J1030 field, with a similar procedure and dataset used in this section. Following this work, we set a threshold of at least five filters, that corresponds to an rms$\approx$0.1 when comparing the $z_{phot}$ with the available spectroscopic redshift in the field.

\subsection{Catalogs} \label{catalogs}
\begin{figure}[t!]
    \begin{center}
        \includegraphics[width=0.5\textwidth]{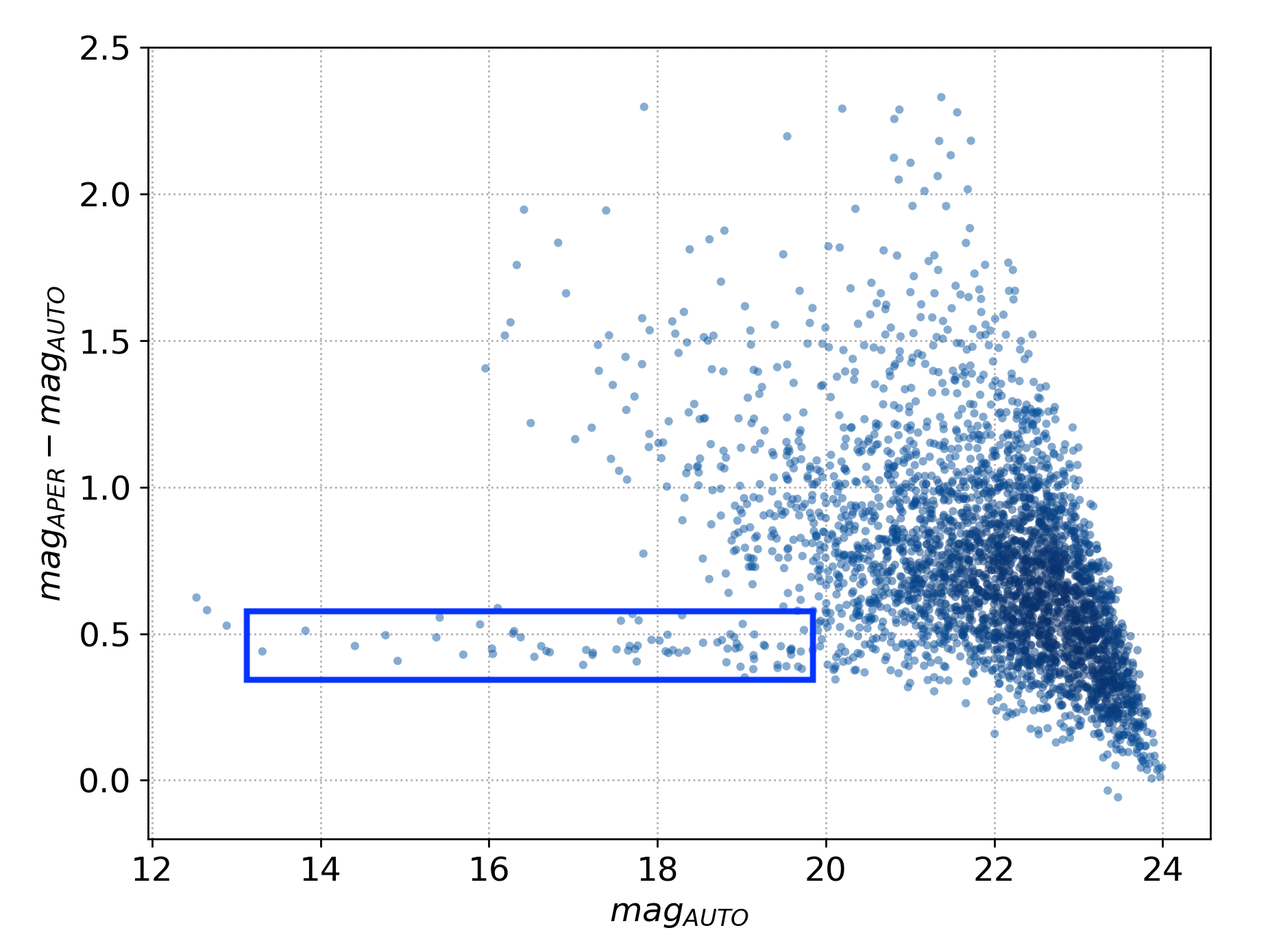}\hfil
        \caption{Distribution of the difference between aperture and total K-band magnitudes in the MUSYC K-deep catalog. The point-like sources are placed in the blue box, where the mean difference between the two magnitudes corresponds to the seeing aperture-correction term to apply.}
    \label{kdeep_corr}
    \end{center}
\end{figure}
In order to correctly reproduce the SED of each individual source, photometry should sample the same physical region of the host galaxy in all bands. Therefore, the photometric redshift technique uses  aperture-corrected magnitudes to account for the flux lost outside the fixed aperture, due to the different seeing conditions in different observations. The photometry in the ONIR catalogs (LBT, WIRCam and MUSYCs) was obtained assuming a circular aperture of diameter $\sim 1.6\arcsec$ (\citealt{gawiser06, quadri07, blanc08, morselli14, balmaverde17}). While in the LBC, WIRCam and BVR MUSYC catalogs the aperture correction was already considered, we estimated the aperture-correction terms in the MUSYC K-deep and K-wide catalogs as follows. 
For point-like sources, the difference between total and aperture magnitudes is, by definition, the aperture correction.
The mag$_{\mathrm{AUTO}}$ entries in the MUSYC K-deep and K-wide catalogs are a good approximation of the total magnitudes, and in Figure \ref{kdeep_corr} we plotted the difference between them and the aperture magnitudes, against the total magnitudes. It is evident that point-like sources are arranged along a straight line, whose value corresponds to the aperture-correction term to be applied to the catalog. We estimated and applied a correction of 0.45 and 0.62 for the MUSYC K-deep and K-wide catalogs, respectively.
All the ONIR catalogs were then matched together assuming a matching radius of 1$\arcsec$ and the resulting sources were then associated with the X-ray counterparts through a likelihood ratio (\citealt{sutherland92}) algorithm\footnote{{\url{https://github.com/alessandropeca/LYR\_PythonLikelihoodRatio}}}, as widely discussed in \citetalias{nanni20}.

The IRAC channels 1 and 2, respectively at 3.6 and 4.5 $\mu m$, were introduced to improve the $z_{phot}$ estimate. 
The inclusion of photometric points at longer wavelengths may increase the best-fit quality but, due to the low angular resolution of the IRAC camera, separating the emission of close sources was not always possible. Therefore, we performed a visual check to associate the correct MIR counterpart to each X-ray source and, to avoid any contamination from blended sources, we excluded ambiguous cases from the SEDs fitting procedure.
Unlike the ONIR catalogs, whose aperture extraction diameters are similar, the IRAC fluxes are provided with aperture diameters of 3.8$\arcsec$ and 5.8$\arcsec$. We decided to use fluxes with the smaller aperture, to minimize blending effects.
To take into account the larger aperture and the different angular resolution, we built the SEDs adding in quadrature $\Delta \mathrm{mag} = 0.1$ to the IRAC magnitude error of each source, in both channel 1 and channel 2.

\section{Results}\label{ch5}
\subsection{X-ray and photometric redshift solutions} \label{zvsz}

The main sample contains 54 X-ray selected obscured AGN candidates with $\approx 80$ median extracted net counts. Each source has been extensively analyzed in the X-rays and through SED fitting in the ONIR and MIR bands. To verify the goodness and the quality of the X-ray redshift solutions, these are now compared with the obtained photometric redshifts.
It was possible to estimate an X-ray redshift for 38 ($\sim 70\%$) sources, down to $\sim$30 net counts, and a photometric redshift for 46 ($\sim85\%$) sources.  We do not report X-ray redshift solutions for sources without significant spectral features, nor when we classified them as unreliable according to the spectral simulations, as discussed in Section \ref{ch3}. Sources without a photometric redshift estimate are those detected in less than five filters.
We obtained both $z_X$ and $z_{phot}$ solutions for 33 ($\sim 61\%$) sources, and for all but three there is a redshift estimate from at least one method. For XID 29, 130 and 135 no redshift could be found by any method.
XID 130 and 135 are very faint X-ray sources (only 26 and 39 full band net counts were extracted, respectively) and lie at very large off-axis angles (10.0$\arcmin$ and 6.9$\arcmin$, respectively). Because of the poor spectral quality, no significant features were detected in their X-ray spectra. XID 29 has 133 net counts in the full band and lies at 2.9$\arcmin$, but no significant features were found in the X-ray spectrum.
XID 130 is detected only in IRAC, while XID 29 and 135 do not reach the threshold on the number of filters, preventing the photometric redshift estimate.
All the obtained redshift solutions are summarized in Table \ref{my_results}, and they represent the first list of redshifts for the J1030 field.

\begin{figure}[t!]
	\begin{center}
		\includegraphics[width=0.5\textwidth]{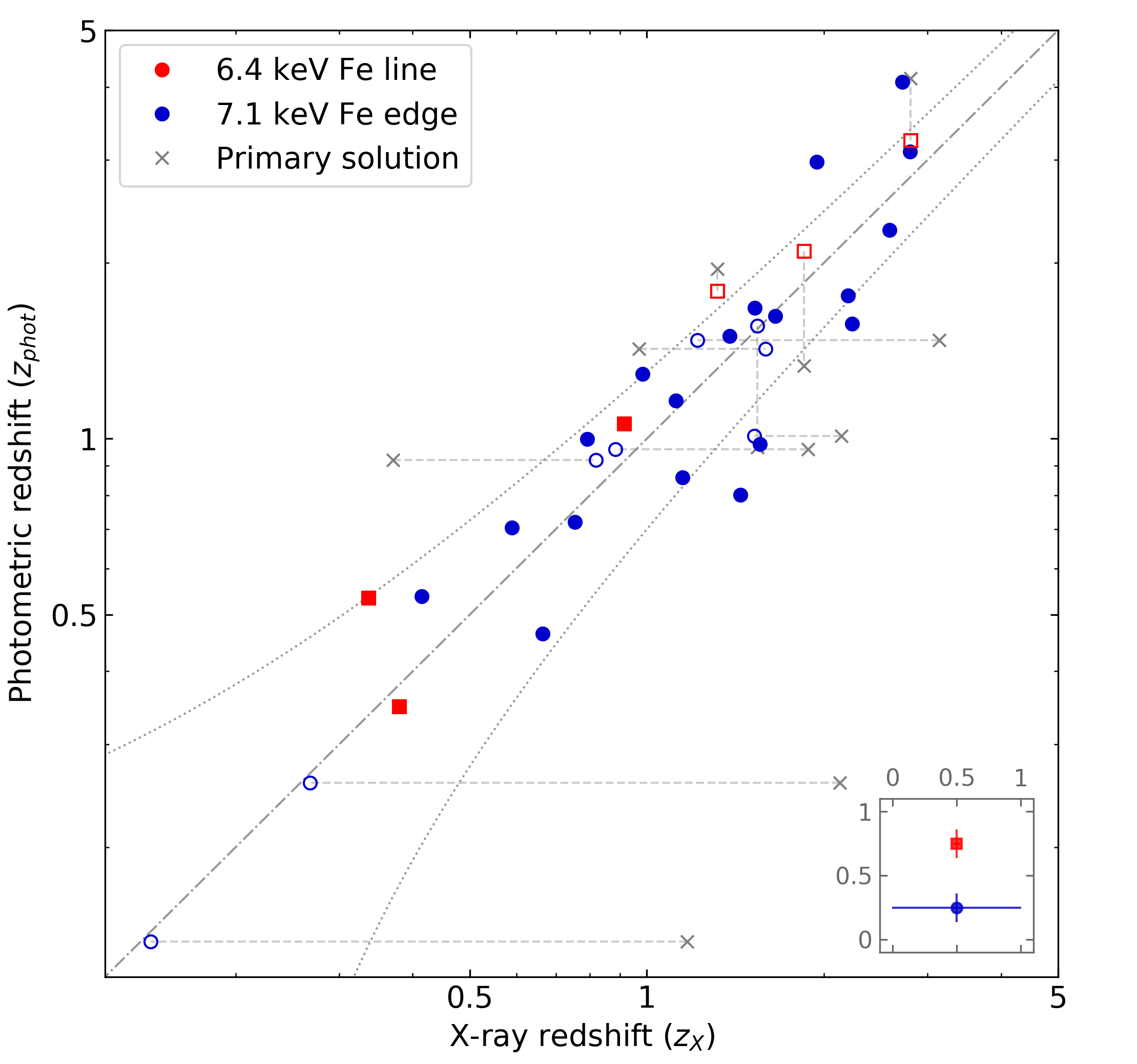}\hfil
		\caption{Comparison between X-ray and photometric redshifts solutions. The red squares indicate the sources in which the 6.4 keV Fe K$\alpha$ line was detected, while the blue dots refer to X-ray solutions identified only with absorption features. The empty markers are the secondary solutions constrained by one of the two methods: if a photometric (X-ray) secondary solution is constrained by a single and unique X-ray (photometric) solution, then the shift between the primary photometric (X-ray) solution (grey crosses) and the constrained solution (empty markers) is indicated with a vertical (horizontal) dashed segment. The gray dotted lines indicate the chosen $z_{phot}$=$\pm 0.15(1+z_{phot})$ confidence region (see text for details), while the point-dotted grey line is the one to one relation. In the lower-right corner we show the average $1\sigma$ errors for solutions with and without the 6.4 keV Fe K$\alpha$ line, in linear scale and for generic redshifts in both axis.}
		\label{final_plot}
	\end{center}
\end{figure}

In Figure \ref{final_plot} we show the comparison between the derived X-ray and photometric redshifts.
The likelihood profiles of $z_X$ and $z_{phot}$ do not always provide a unique solution. In fact, especially for faint sources where the data quality is relatively poor, both distributions may have non-negligible secondary solutions.
If a primary X-ray solution matches (i.e., it is consistent within the errors) with a primary photometric solution (22/33 cases), we discard any secondary solutions. However, the remaining 11 cases ($\sim20\%$ of the main sample) have secondary solutions, obtained with one method, that match with the primary and unique solutions derived with the other. In these cases the primary solution from one method can constrain secondary solutions obtained with the other, hence we selected the agreed redshift as the unique solution (empty markers). There are no cases with a clear match between secondary X-ray and photometric solutions.
The blue dots refer to X-ray solutions driven only by absorption features, while the red squares indicate sources in which the 6.4 keV Fe K$\alpha$ line has been detected. The majority ($\sim82\%$) of the solutions are driven by absorption features.
As shown in the inset of the same figure, the uncertainties of the X-ray redshift solutions are larger for those sources in which it was not possible to identify any clear Fe K$\alpha$ emission line. This is explained by the fact that the Fe K$\alpha$ emission line is a very narrow feature compared to the Fe absorption edge. Thus, in case of a detected Fe K$\alpha$ line, the X-ray redshift probability sharply decreases before and after the best fit value, resulting in a smaller uncertainty.

\begin{figure*}[!t]
	\begin{center}
		\includegraphics[width=1.0\textwidth]{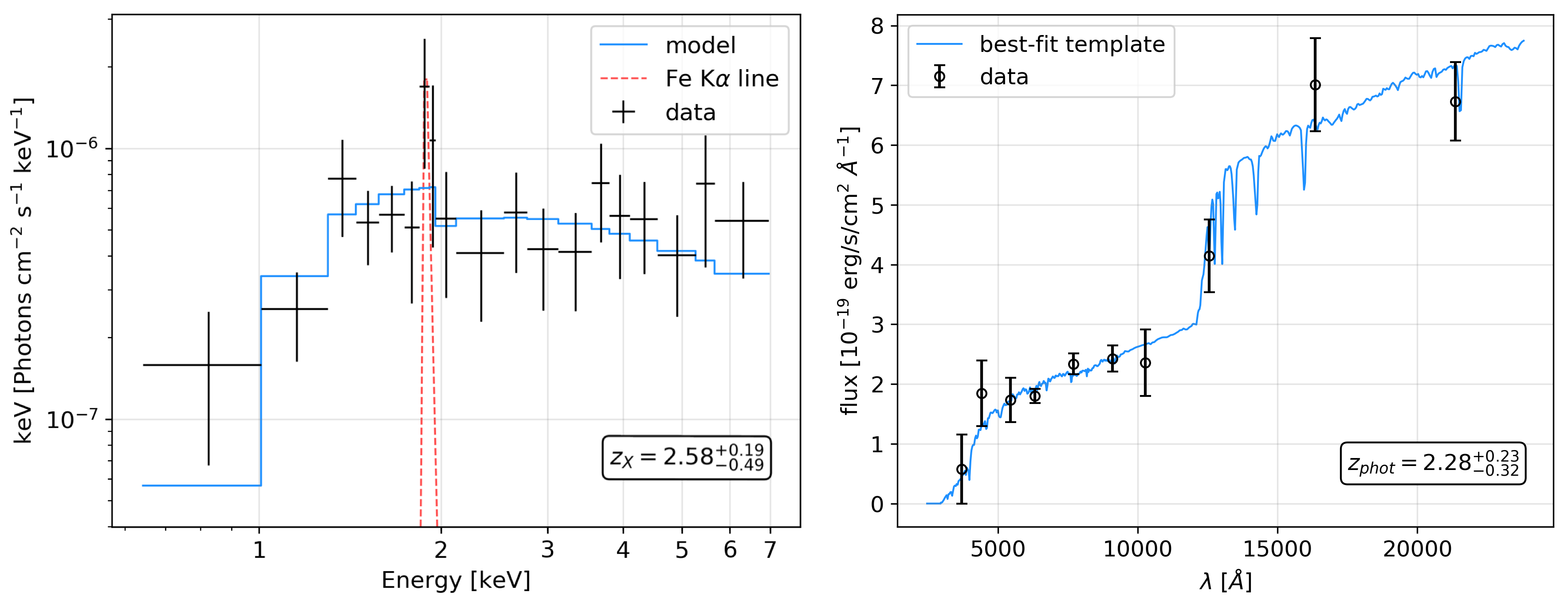}\hfil
		\caption{Example of a sample source (XID 41, 118 full band net counts) with spectroscopic redshift ($z_{spec}=2.511$, \citealt{kriek08}) in agreement with those obtained by the X-ray spectral analisys and SED fitting. Left panel: the X-ray spectrum (black points), rebinned for graphic purposes, with its best-fit model (blue solid line). A prominent Fe 7.1 keV (rest-frame) absorption edge at $\sim$2 keV (observed-frame) and a photoelectric absorption cut-off at softer energies are evident. These features are produced by a column density $N_H=1.7_{-0.4}^{+0.5} \times 10^{23}$ cm$^{-2}$ and drive the X-ray redshift solution. The Fe K$\alpha$ emission line (red dashed curve), is only tentatively detected (1$\sigma$) but not fitted. Right panel: observed SED (black points) and best-fit template (blue solid line), corresponding to a red and passive galaxy where the 4000 $\AA$ break is clearly identified at $\sim 13000$ $\AA$ (observed-frame).}
		\label{confirm_src}
	\end{center}
\end{figure*}
Overall, there is a good correlation between $z_X$ and $z_{phot}$ despite a non-negligible scatter.
Considering a typical accuracy\footnote{Defined as $\mathrm{rms}=\langle \frac{|z_i - z_j|}{1+z_j} \rangle$, where ($z_i$,$z_j$) are ($z_{phot}$,$z_{spec}$) for photometric redshifts, ($z_X$,$z_{sim}$) for X-ray redshifts, and ($z_X$,$z_{phot}$) in our results.} of rms=0.1 for photometric redshifts obtained with a similar number of filters (e.g., \citealt{capak04, zheng04}, Marchesi et al. in prep.), and rms=0.1 for the X-ray simulations described in Section \ref{semitheor_sim}, we assumed a confidence region of $\pm0.15(1+z)$, where 0.15 was computed by summing in quadrature the two rms terms.
About 76$\%$ of the sources fall within the chosen confidence region (grey dotted lines).
We defined as outliers those sources whose $z_{phot}$ value does not lie within the $z_X$ 1$\sigma$ error bars, and the $z_X$ value is outside the 0.15(1+$z_{phot}$) confidence region. The outlier fraction is then 9\% (3/33 sources), and the total fraction of sources where we constrained a redshift solution with both methods is 56\% (30/54). We achieved an accuracy of rms=0.10 and, when considering only the primary solutions with both methods, the rms increases to $\approx$0.2.
These rms values and the outlier fraction are comparable to those obtained in other X-ray redshift techniques. For instance, \citet{simmonds18} obtained a rms$\approx$0.2 and 8\% outlier fraction when validating against reliable spectroscopic redshifts. This indicates that the assumptions adopted in our procedure can be considered appropriate.

We show one of the sample sources (XID 41) in Figure \ref{confirm_src}, for which there is a spectroscopic redshift measurement from \cite{kriek08}, $z_{spec} = 2.511$. This is the only spectroscopic redshift available so far for our sample.
On the one side, we derived an X-ray redshift solution, $z_X=2.58_{-0.49}^{+0.19}$, by fitting a prominent Fe 7.1 keV edge coupled with the photoelectric absorption cut-off. These features are produced by a heavily obscured, yet Compton-thin AGN with $N_H=1.7_{-0.4}^{+0.5} \times 10^{23}$ cm$^{-2}$ (left panel).
On the other side, the derived photometric redshift solution, $z_{phot}=2.28_{-0.32}^{+0.23}$, is driven by a strong drop in the SED identified at $\sim12500 \AA$ (right panel). This feature can be associated with a prominent 4000 $\AA$ break (e.g., \citealt{bruzual83}, \citealt{kauffmann03}), which indicates that the host is a red and passive galaxy. Both our solutions are consistent, within the uncertainties, with the spectroscopic redshift.

\subsection{Sample properties} \label{agn_anal}
\begin{figure}[t!]
	\begin{center}
		\includegraphics[width=0.5\textwidth]{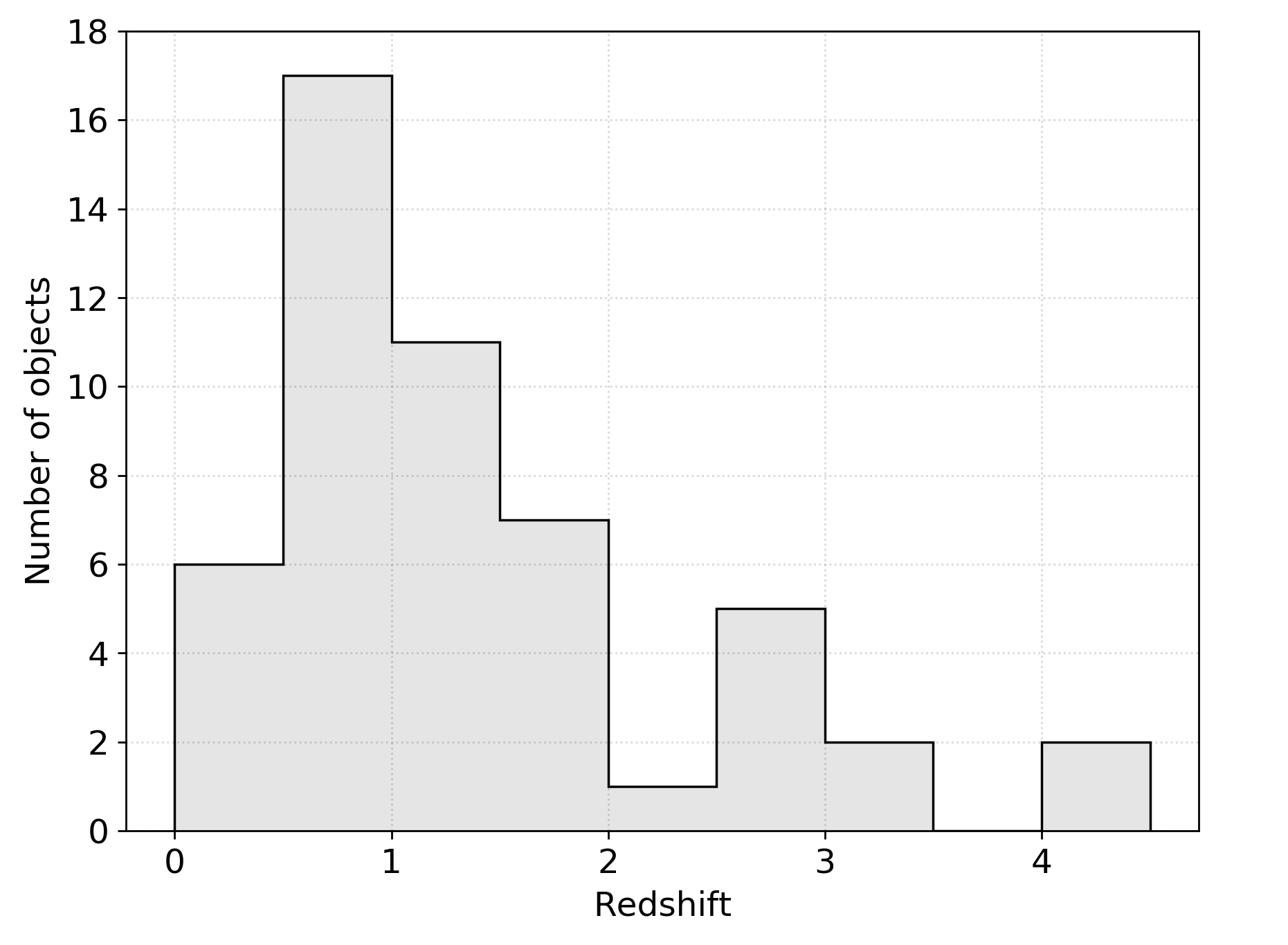}\vspace{0.07cm}
		\includegraphics[width=0.5\textwidth]{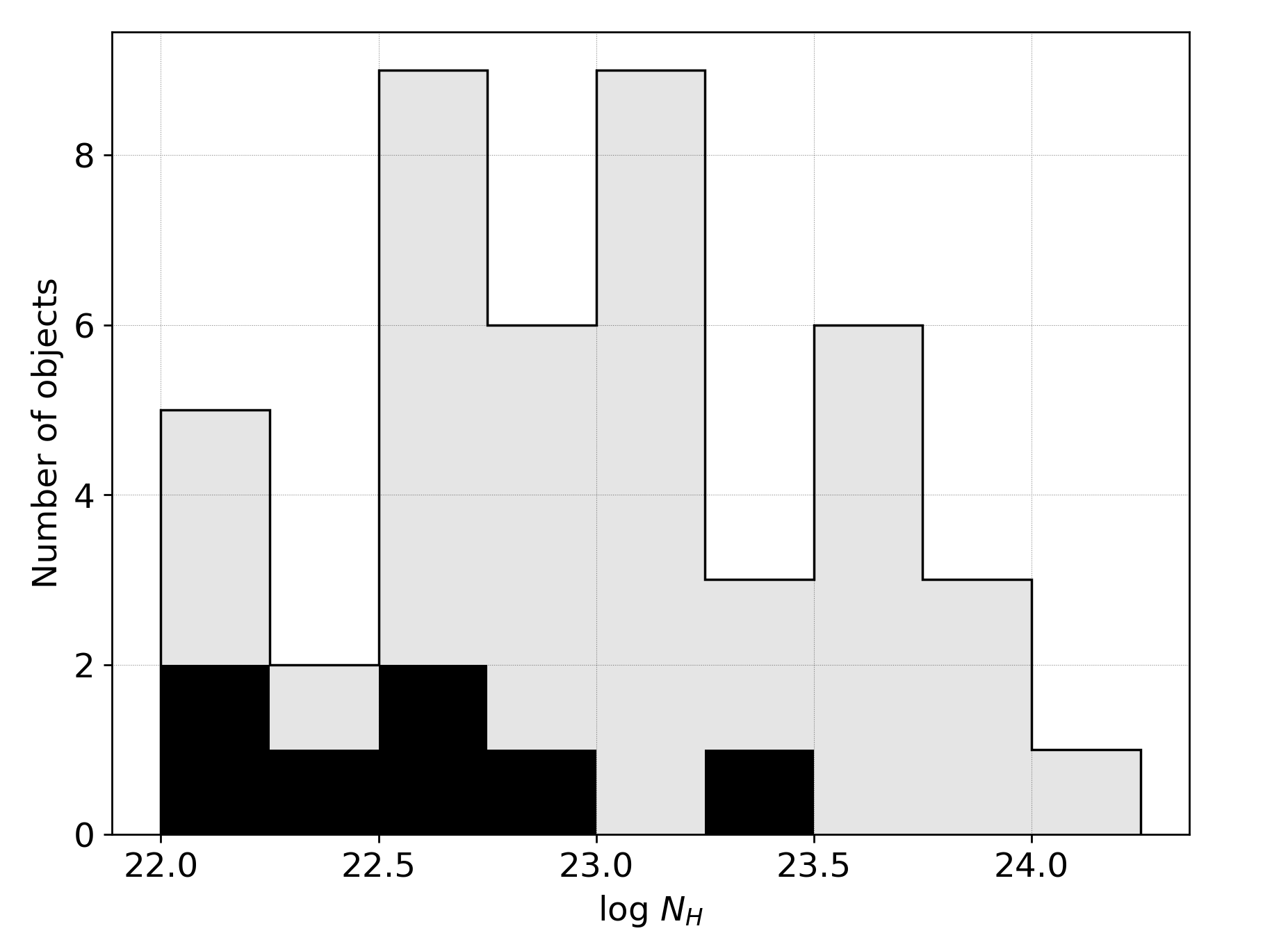}\vspace{0.07cm}
		\includegraphics[width=0.5\textwidth]{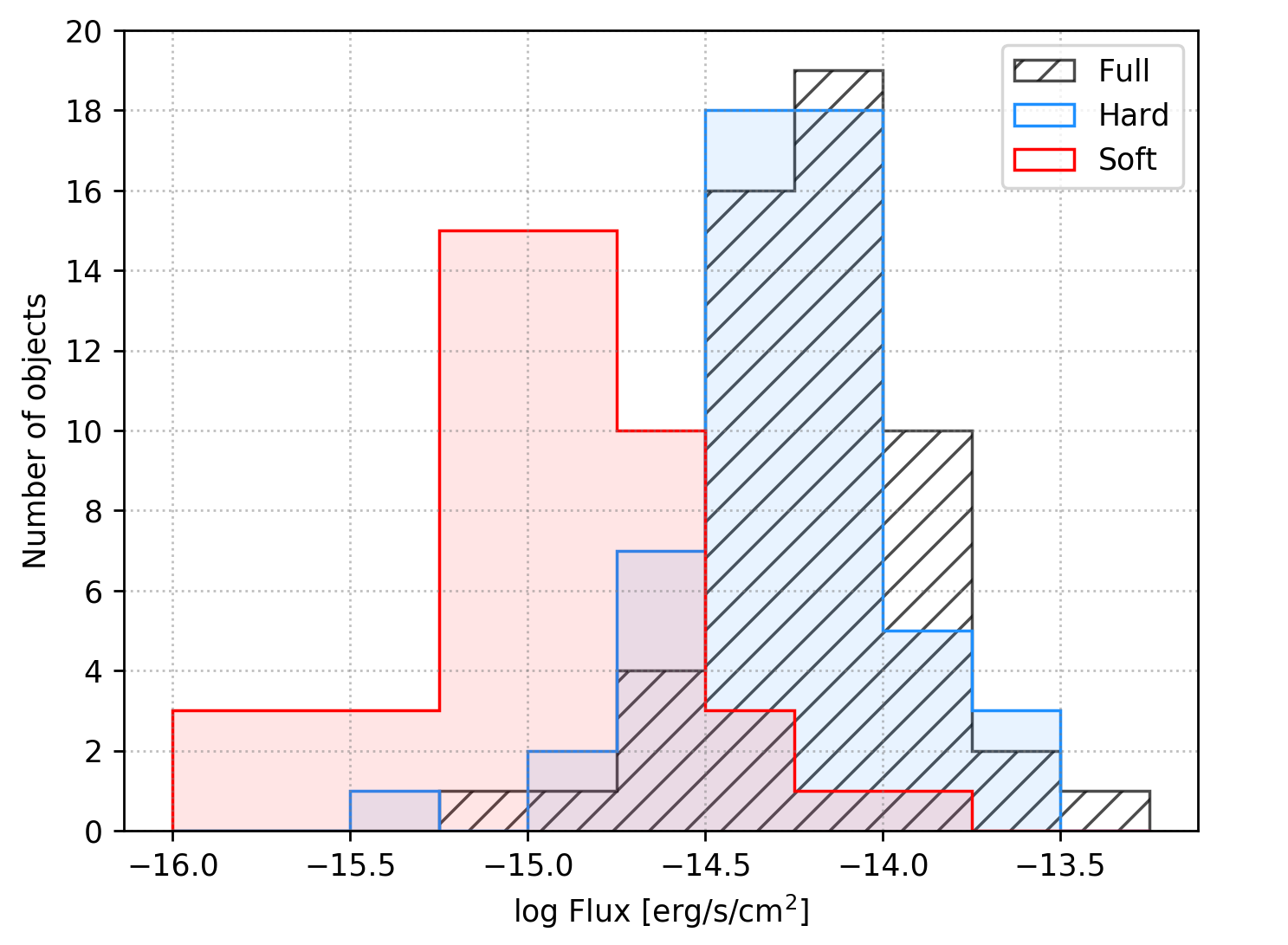}
		\caption{Redshift (top panel) and column density (middle panel) distributions derived from our analysis. Column density upper limits are plotted in full black. In the bottom panel the observed flux distributions in the soft (red), hard (light blue) and full (hatched black) bands are shown.}
		\label{z_nh_distrib}
	\end{center}
\end{figure}

Given that the good agreement between the two methods validates the obtained redshift solutions, it was possible to conduct a study on the physical and intrinsic X-ray properties of the selected obscured AGN sample. Photometric redshifts generally have smaller uncertainties than the X-ray ones, except for those cases where the Fe K$\alpha$ line was detected. We therefore decided to use $z_X$ when the Fe K$\alpha$ line was identified, and $z_{phot}$ elsewhere. When none of the two above solutions was available, we used absorption driven $z_X$, if estimated. The analysis was then feasible for 51 sources.
The redshift distribution (Figure \ref{z_nh_distrib}, top panel), spans from $\sim$0.1 to $\sim$4 with a median value of $z=1.3$, and is peaked between 0.5 and 1 in agreement with the obscured redshift distributions in other deep X-ray surveys (e.g., \citealt{fiore08, georg08}).

We performed the spectral analysis adopting an absorbed power-law model, as described in Section \ref{xray_spec}, but now fixing the redshift.
We also carried out a few tests by fitting independently the source and background spectrum using a background model (see \citealt{marchesi16b} for details), and did not find significant differences in the derived best-fit source parameters.

\begin{figure}[t!]
	\begin{center}
		\includegraphics[width=0.5\textwidth]{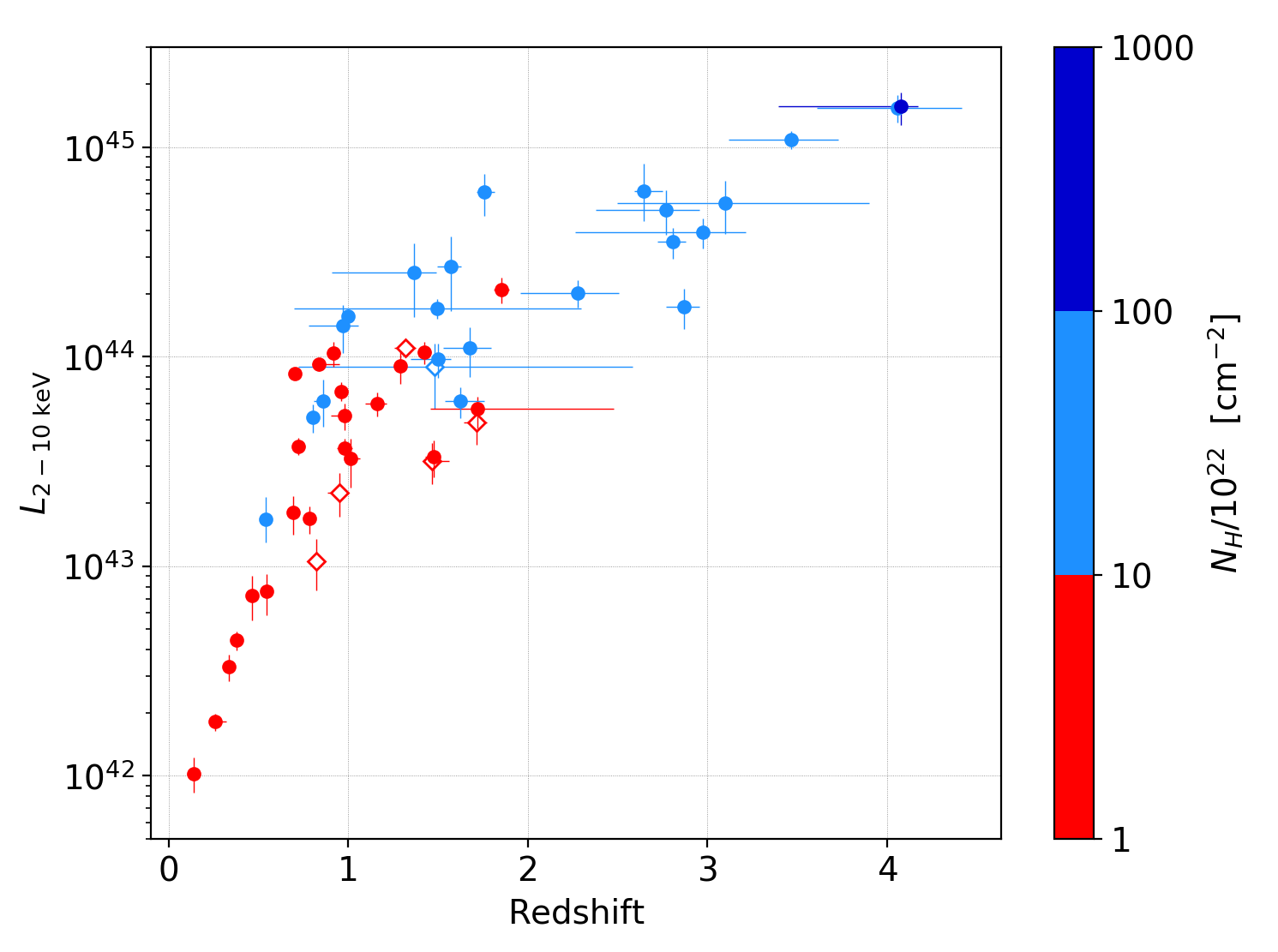}\hfil
		\caption{Intrinsic, absorption-corrected luminosity in the 2-10 keV rest-frame band, as a function of redshift and $N_H$ in color code. Empty diamonds represent $N_H$ upper limits.}
		\label{deabs_lum}
	\end{center}
\end{figure}

The derived column density distribution (Figure \ref{z_nh_distrib}, middle panel) ranges between $10^{22}$ and $10^{24}$ cm$^{-2}$, with a mean value of $N_H= 1.7 \times 10^{23}$ cm$^{-2}$ typical of Compton-thin AGN, plus one Compton-thick AGN candidate ($N_H \sim 1.1 \times 10^{24}$ cm$^{-2}$) at $z\approx 4$.
We show the obtained flux distributions in the full, soft and hard bands in Figure \ref{z_nh_distrib} (bottom panel).
The derived intrinsic, absorption-corrected, 2-10 keV rest-frame X-ray luminosity is in the range $\sim 10^{42}-10^{45}$ erg s$^{-1}$, with a median value of $L_{\mathrm{2-10\, keV}} = 8.3\times10^{43}$ erg s$^{-1}$.
We report also a very low ($\sim10^{40}$ erg s$^{-1}$) luminosity object (XID 127), whose counterpart is extended in the optical images. It is identified as a very bright nearby galaxy ($z_{phot}=0.05$, $R=18$), and its X-ray emission may be produced by a very low luminosity AGN or stellar processes.
Overall, we found that the level of obscuration increases as both redshift and luminosity increase (Figure \ref{deabs_lum}). In particular, at $L_{2-10 keV}<10^{44}$ erg s$^{-1}$ and $z>1.5$, AGN with $N_H>10^{23}$ cm$^{-2}$ start dominating over AGN with column densities between $10^{22}$ and $10^{23}$ cm$^{-2}$.
However, we are biased towards the most obscured and luminous objects as the redshift increases. On the one hand, we lose $N_H<10^{23}$ cm$^{-2}$ sources at high-redshift ($z\gtrsim 3$) because of the $HR$ selection (see Figure \ref{tozzi_HR}). In addition, since the X-ray photoelectric absorption cut-off moves towards the lower boundary of the observational band (0.5 keV), the estimate of low $N_H$ values becomes difficult (e.g., \citealt{tozzi06, marchesi16b}).
On the other hand, we only see the brightest sources because the sample is flux limited.
We also point out that a fraction of obscured AGN at lower net counts regimes (Marchesi et. al. in prep) is probably missing.
Our results are summarized in Table \ref{my_results}.

\subsection{Overdensity AGN candidates}
In \citet{gilli19} we reported the discovery of a galaxy overdensity at $z \approx 1.7$. The structure has eight members confirmed by secure ONIR spectroscopic redshifts (VLT/MUSE and LBT/LUCI, \citealt{gilli19}), plus three members confirmed by ALMA spectra (\citealt{damato20}). Ten of them are star-forming galaxies and one, located at the center of the overdensity, is a Compton-thick ($N_H=1.5_{-0.5}^{+0.6} \times 10^{24}$ cm$^{-2}$) Fanaroff–Riley type II (FRII) radio galaxy. This source is the only one detected in the X-rays (XID 189 in \citetalias{nanni20}).
Due to the limited area ($\sim$1-1.5 arcmin) covered around the FRII core by ONIR spectroscopic and ALMA observations, we were able to estimate an overdensity projected size of at least $\sim$800 kpc.
In this work, we found six sources (XID 37, 40, 48, 69, 131, and 137) whose redshift solutions are consistent with $z=1.7$ and, therefore, may be new overdensity members. In particular, XID 37 and 131 have both photometric and X-ray redshift estimates, while for XID 69 and the remaining sources we derived only the X-ray and photometric redshift solutions, respectively.
Using an angular scale of 8.5 kpc arcsec$^{-1}$, valid for the adopted cosmology at $z=1.7$, the projected distances between these sources and the FRII core are in the range 2-4 Mpc, suggesting that the structure may be more extended than previously estimated. 
The proposed technique can hence be used to identify possible AGN members of cosmological structures, such as galaxy overdensities and proto-clusters, which may be extended up to several Mpc (e.g., \citealt{gilli03, overzier16}).

\section{Discussion}\label{ch6}

\subsection{Reliability of the obscured AGN selection from HR analysis}\label{ch61}
\begin{figure}[t!]
	\begin{center}
		\includegraphics[width=0.5\textwidth]{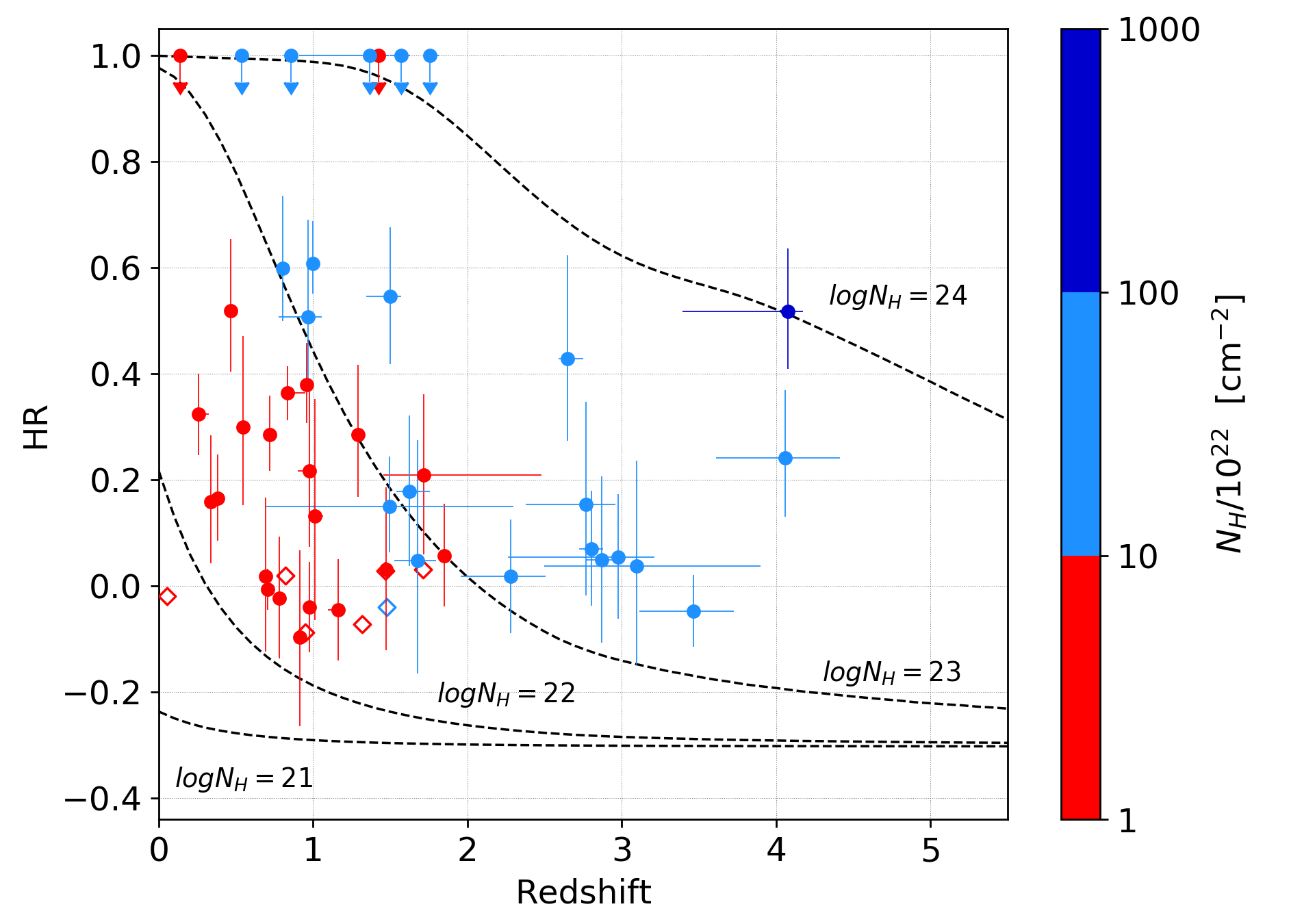}\hfil
		\caption{Redshift and column densities (in color code) obtained from our analysis, plotted over the expected hardness ratio trends (dashed black lines), computed as discussed in Section \ref{sample_selection} for $\Gamma=1.9$ (see also Figure \ref{tozzi_HR}). Empty diamonds represents $N_H$ upper limits.}
		\label{break_deg}
	\end{center}
\end{figure}

In Figure \ref{break_deg} we compare the observed hardness ratios as a function of the the derived redshifts and column densities, with the expected simulated trends. There is a clear distinction between red ($N_H < 10^{23}$ cm$^{-2}$) and light blue points ($10^{23} < N_H < 10^{24}$ cm$^{-2}$). The only candidate at $N_H > 10^{24}$ cm$^{-2}$ (dark blue point) also lies in the corresponding region. This confirms that a hardness ratio threshold, carefully calibrated on the data, can be a good proxy for the selection of obscured AGN, as discussed before. However, other than a dependence on the instrument effective area, the expected $HR$ is also a function of the AGN spectral shape, as described in Section \ref{sample_selection} for typical $\Gamma$ values.

In the following, we discuss how the $HR$ trends may be influenced by commonly observed AGN spectral features:  soft excess and reflection.
The soft excess is possibly produced by scattered radiation into the line of sight (e.g., \citealt{ueda07,brightman12}), and is emitted over the primary AGN continuum at rest-frame energies below $\sim$1$-$2 keV.
Especially for high-obscuration levels ($N_H>10^{23}$ cm$^{-2}$), where the main power-law radiation is strongly extinguished at soft rest-frame energies, the soft excess may give a non-negligible contribution to the observed soft band, decreasing the $HR$ value for sources up to $z\sim$2$-$3.
However, its emission is observed to be $<10\%$ of the primary component (e.g., \citealt{lanzuisi15}), with typical values of 1-3 \% (e.g., \citealt{gilli07,ricci17}), and its contribution to the observed $HR$ is diluted because of the strong decrease of the \textit{Chandra} collecting efficiency at $E\lesssim$1 keV. The combination of these factors keeps the $HR=-0.1$ threshold valid (see Appendix \ref{app_B1}). 
The reflected emission is produced by the reprocessing of the primary X-ray continuum by circumnuclear material, and it may also influence the $HR$ for sources with high $N_H$. In general, the reflection component peaks at rest-frame energies of $\sim30$ keV (e.g., \citealt{ajello08}), with a small to moderate contribution in the observed soft band even at redshift $\sim$3$-$4. As a consequence, we do not expect to miss obscured AGN with the adopted $HR$ selection criteria.
A detailed modelling of these spectral components is shown in Appendix \ref{app_B}.

\subsection{Applicability of the method}
In the following we discuss the main differences and similarities between our procedure and other X-ray redshift techniques adopted in AGN surveys (e.g., \citealt{civano05,iwasawa12, simmonds18, iwasawa20}):
\begin{enumerate}[wide,nosep,label=(\roman*)]
\item We constrained X-ray redshifts down to a lower net counts regime, $\sim 30$ net counts for particular cases (see Section \ref{semitheor_sim}). This is similar to \citet{simmonds18}, but markedly different to the hundreds of counts required in other K$\alpha$-based redshift derivations.
Because the procedure has strong dependencies on the number of net counts, and considering that it is calibrated on the instrumental response, we expect this net counts threshold to be a reliable lower limit for other \textit{Chandra} deep observations.
\item The significance of each X-ray redshift solutions is analyzed through extensive simulations based on both local (Section \ref{lin_sim}) and, differently from any other method, global (Section \ref{semitheor_sim}) backgrounds and responses. For the latter, we directly extracted the background as a function of the off-axis angle, to get a realistic background estimate. Then, we re-scaled and associated it to the simulated source spectra by taking into account the PSF size and the instrumental response at the source position. Taking into account the off-axis angle effects is crucial to maximize and properly interpret the X-ray information, especially for pointed fields. A different solution for the background handling is to model it. For example, \citet{simmonds18} sampled a single, representative background for each instrument using archival observations. We refer to that article for additional information, and to \citet{buchner14} for details about background modelling in X-ray data.
\item As discussed in Section \ref{zvsz}, there is a non-negligible fraction ($\sim20\%$) of sources in which a secondary redshift solution was constrained by a unique $z_{phot}$ or $z_X$. This suggests that the combined use of photometric and X-ray solutions can solve redshift degeneracies that often arise in both methods, and may provide better redshift estimates (see e.g., \citealt{vignali15,simmonds18,iwasawa20}). For this reason, photometric redshifts are included in the procedure and, for these particular cases, the final redshift solutions are driven by the combination of the two methods.
\end{enumerate}

Considering the above points, the proposed procedure can be used in current X-ray deep fields, where there is still a fraction ($\sim$5\%) of X-ray sources without a solid redshift estimate (e.g., \citealt{salvato09, marchesi16a}). 
For example, taking sources with at least 50 net counts and $HR>-0.1$ in the 7 Ms CDF-S (\citealt{luo17}) and the AEGIS-XD (\citealt{nandra15}) surveys, the fraction of objects without any redshift estimate is 1-2\% (11 in both fields), and there is a 3-4\% of sources (37 and 26, respectively) associated only with photometric redshifts that have non-negligible secondary solutions. 
Moreover, $11\%$ of sources in the CDF-S are single-line spectroscopic redshifts (e.g., \citealt{szokoly04}), which are often not reliable (e.g., \citealt{simmonds18, barger19}), and the AEGIS-XD survey has $12\%$ of sources marked with a poor quality flag.
These cases are optimal for the proposed procedure, which may estimate reliable redshifts from the X-ray spectra alone and/or compare them with the available redshifts, to better evaluate secondary photometric solutions and single-line spectroscopic redshift.
These sources are potentially high-redshift obscured AGN, which are very challenging to be revealed with the current methods (e.g., \citealt{cowie20}) and may contribute to the high-redshift obscured AGN fraction, which is still uncertain and subject of debate (\citealt{georgakakis17,vito18}).

We applied the X-ray redshift procedure to the J1030 field, which lacks the massive spectroscopic coverage of the other X-ray deep fields.
\citet{simmonds18} showed through simulations that X-ray redshifts may be derived from observations of existing missions (\textit{Chandra}, XMM-\textit{Newton}, \textit{NuSTAR}, \textit{Swift/XRT}). Similarly, the proposed method can be applied to future deep \textit{Chandra} and XMM-\textit{Newton} observations, to the forthcoming \textit{eROSITA} eRASS surveys (e.g., \citealt{merloni13}), to the future \textit{Athena} observatory (e.g., \citealt{aird13}) and, hopefully, to missions under study such as \textit{Lynx} (e.g., \citealt{gaskin19}) or \textit{AXIS} (e.g., \citealt{mushotzky19}), which will observe faint heavily obscured objects with no or extremely faint ONIR counterparts.
While \textit{eROSITA}'s soft response and short exposures limit the applicability of X-ray redshift methods (\citealt{simmonds18}), large-area, high-resolution missions such as \textit{Athena} will likely benefit tremendously from X-ray redshifts.
In deep \textit{Athena}, \textit{Lynx} and \textit{AXIS} surveys, obtaining adequately deep photometric data and/or ONIR spectroscopy will be very costly.
Thus, by applying a method similar to the one described in this work, and tuning the $HR$ and number of counts thresholds to take into account the effective area of future missions, it will be possible to provide reliable X-ray redshift solutions for obscured AGN.

\section{Summary}\label{ch7}

We proposed a multi-wavelength method to constrain the redshifts of X-ray selected obscured AGN, and applied it to the analysis of a sample of 54 candidates in the field around the $z=6.3$ QSO SDSS J1030+0524. The described technique involves X-ray photometry, spectral analysis and spectral simulations applied to the \textit{Chandra} $\sim$479 ks observational campaign, combined with a SED fitting procedure that includes ONIR and MIR photometry from LBT/LBC ($r$, $i$, $z$ bands), CFHT/WIRCam ($Y$, $J$ band), as well as the MUSYC BVR, K-wide and K-deep catalogs ($U,B,V,R,I,z,J,H,K$ bands), and \textit{Spitzer}/IRAC channel 1 and 2 at 3.6 and 4.5 $\mu m$, respectively. 

\noindent Our main results are summarized as follows:
\begin{itemize}[wide,nosep]
    \item We derived reliable X-ray redshifts for a sample of obscured AGN candidates with hardness ratio $HR>-0.1$. We selected sources in \citetalias{nanni20} detected with at least 50 full band net counts, so as to identify the main X-ray spectral features like the Fe K$\alpha$ 6.4 keV emission line and the Fe 7.1 keV absorption edge. The identified features were then validated through ad-hoc spectral simulations.
    \item We computed photometric redshifts trough a SED fitting procedure to validate the derived X-ray solutions. The comparison between $z_X$ and $z_{phot}$ gave an accuracy of rms=0.10, and revealed that the combined use of both methods can constrain secondary solutions from both sides. We obtained a reliable redshift solution with at least one method for 51 ($\sim 94\%$) sources, with a median value of $z=1.3$ in the range $z\sim 0.1-4$.
    \item The obtained redshift solutions were used to derive the X-ray physical intrinsic properties of the sample. We derived a mainly Compton-thin AGN population ($10^{22}\lesssim N_H \lesssim 10^{24}$ cm$^{-2}$) with a median value of $N_H=1.7\times 10^{23}$ cm$^{-2}$, and intrinsic, absorption-corrected, rest frame 2-10 keV luminosities in the range $10^{42}-10^{45}$ erg s$^{-1}$ with a median $L_{\mathrm{2-10\, keV}} = 8.3\times10^{43}$ erg s$^{-1}$, similar to the distributions observed for obscured AGN in other deep X-ray surveys.
    \item We found six possible new AGN members of a galaxy overdensity at $z\approx 1.7$, showing that it may be extended up to 4 Mpc and that the proposed technique can be used to find candidates of cosmological structures.
    \item Finally, we discussed the peculiarities and the feasibility of the proposed method in the context of current X-ray deep surveys, where there is still a fraction of X-ray AGN without a solid redshift estimate, as well as in the context of future X-ray observations with both current and planned X-ray facilities.
\end{itemize}

\acknowledgments{
\noindent We acknowledge the anonymous referee for the useful comments that improved the quality of the paper. AP acknowledges D. Costanzo for all the support during the years, and Q. D'Amato for the helpful discussions not only about AGN.
We acknowledge financial contribution from the agreement ASI-INAF n. 2017-14-H.O. NC and AP kindly acknowledge NASA-ADAP grant.
}

%\clearpage
\newpage

\startlongtable
\begin{deluxetable*}{ccccccccc}

\tabletypesize{\normalsize}
\tablecaption{Properties of obscured AGN candidates ($HR>-0.1$ and net counts $\geq 50$ from \citetalias{nanni20}) in the J1030 field.}
\tablehead{
\colhead{XID} & \colhead{Cts full} & \colhead{HR} & \colhead{$z_{phot}$} & \colhead{$z_{X}$} &\colhead{$N_H$}& \colhead{$F_X$} & \colhead{$L_X$} & \colhead{$feat.$}\\
(1) & (2) & (3) & (4) & (5) & (6) & (7) & (8) & (9)
}

%\hline
\startdata
2           & $890         _{-30          }^{+31          }$ & $-0.01       _{-0.04        }^{+0.04        }$ & $0.70        _{-0.02        }^{+0.02        }$ & $0.59        _{-0.36        }^{+0.12        }$ & $1.6         _{-0.2         }^{+0.2         }$ & $28.39       _{-1.35        }^{+1.25        }$ & $  8.31 _{- 0.39 }^{+ 0.39 } $ & edge        \Tstrut\Bstrut\\ 
3           & $141         _{-12          }^{+13          }$ & $-0.04       _{-0.08        }^{+0.08        }$ & $0.98        _{-0.04        }^{+0.04        }$ & $-1                                          $ & $2.4         _{-0.7         }^{+0.8         }$ & $5.49        _{-0.53        }^{+0.52        }$ & $3.65 _{- 0.44 }^{+ 0.43 } $ & -           \Tstrut\Bstrut\\ 
8           & $115         _{-11          }^{+12          }$ & $0.07        _{-0.11        }^{+0.11        }$ & $3.24        _{-0.04        }^{+0.41        }$ & $2.80        _{-0.05        }^{+0.05        }$ & $33.4        _{-9.2         }^{+9.9         }$ & $3.38        _{-0.46        }^{+0.41        }$ & $  35.38 _{- 5.95 }^{+ 5.67 } $ & line        \Tstrut\Bstrut\\ 
11          & $95          _{-10          }^{+11          }$ & $0.05        _{-0.12        }^{+0.12        }$ & $2.98        _{-0.71        }^{+0.24        }$ & $1.94        _{-0.38        }^{+1.05        }$ & $27.4        _{-6.9         }^{+8.0         }$ & $3.67        _{-0.53        }^{+0.40        }$ & $  39.41 _{- 6.57 }^{+ 6.11 } $ & edge        \Tstrut\Bstrut\\ 
15          & $407         _{-20          }^{+21          }$ & $0.36        _{-0.05        }^{+0.05        }$ & $0.94        _{-0.04        }^{+0.11        }$ & $-1                                          $ & $6.4         _{-0.6         }^{+0.6         }$ & $13.64       _{-0.84        }^{+0.82        }$ & $  9.19 _{- 0.63 }^{+ 0.62 } $ & -           \Tstrut\Bstrut\\ 
16          & $55          _{-8           }^{+9           }$ & $0.03        _{-0.15        }^{+0.16        }$ & $1.47        _{-0.05        }^{+0.03        }$ & $1.22        _{-0.74        }^{+0.95        }$ & $4.4         _{-1.9         }^{+2.5         }$ & $1.85        _{-0.34        }^{+0.23        }$ & $  3.33 _{- 0.68 }^{+ 0.65 } $ & edge        \Tstrut\Bstrut\\ 
22          & $65          _{-8           }^{+9           }$ & $0.05        _{-0.16        }^{+0.16        }$ & $-1                                          $ & $2.87        _{-0.06        }^{+0.05        }$ & $32.3        _{-9.7         }^{+12.3        }$ & $1.51        _{-0.27        }^{+0.31        }$ & $  17.36 _{- 3.79 }^{+ 3.73 } $ & line        \Tstrut\Bstrut\\ 
26          & $261         _{-16          }^{+17          }$ & $-0.05       _{-0.07        }^{+0.07        }$ & $-1                                          $ & $3.47        _{-0.21        }^{+0.16        }$ & $23.8        _{-3.8         }^{+6.0         }$ & $7.50        _{-0.62        }^{+0.47        }$ & $  108.52 _{- 10.44 }^{+ 10.28 } $ & edge        \Tstrut\Bstrut\\ 
28          & $374         _{-19          }^{+20          }$ & $-0.07       _{-0.06        }^{+0.06        }$ & $1.79        _{-0.08        }^{+0.08        }$ & $1.32        _{-0.04        }^{+0.04        }$ & $ < 1.6                                     $ & $8.42        _{-0.64        }^{+0.53        }$ & $  10.98 _{- 0.78 }^{+ 0.77 } $ & line        \Tstrut\Bstrut\\ 
29          & $133         _{-16          }^{+17          }$ & $0.42        _{-0.09        }^{+0.09        }$ & $-1                                          $ & $-1                                          $ & $-1                                          $ & $4.57        _{-0.43        }^{+0.53        }$ & $ -1         $ & -           \Tstrut\Bstrut\\ 
30          & $200         _{-14          }^{+15          }$ & $0.32        _{-0.08        }^{+0.08        }$ & $0.26        _{-0.02        }^{+0.06        }$ & $0.27        _{-0.16        }^{+0.70        }$ & $1.9         _{-0.3         }^{+0.3         }$ & $5.95        _{-0.50        }^{+0.48        }$ & $  0.18 _{- 0.02 }^{+ 0.02 } $ & edge        \Tstrut\Bstrut\\ 
36          & $204         _{-15          }^{+16          }$ & $0.38        _{-0.07        }^{+0.08        }$ & $0.96        _{-0.02        }^{+0.04        }$ & $0.88        _{-0.54        }^{+0.35        }$ & $8.1         _{-1.1         }^{+1.2         }$ & $9.53        _{-0.87        }^{+0.65        }$ & $  6.84 _{- 0.71 }^{+ 0.69 } $ & edge        \Tstrut\Bstrut\\ 
37          & $79          _{-9           }^{+10          }$ & $0.18        _{-0.14        }^{+0.14        }$ & $1.62        _{-0.08        }^{+0.14        }$ & $1.65        _{-1.00        }^{+0.43        }$ & $12.3        _{-2.8         }^{+3.4         }$ & $2.46        _{-0.37        }^{+0.34        }$ & $  6.15 _{- 1.07 }^{+ 1.02 } $ & edge        \Tstrut\Bstrut\\ 
38          & $119         _{-11          }^{+12          }$ & $-0.04       _{-0.10        }^{+0.10        }$ & $1.16        _{-0.07        }^{+0.05        }$ & $1.12        _{-0.68        }^{+1.09        }$ & $3.6         _{-1.1         }^{+1.2         }$ & $5.79        _{-0.63        }^{+0.76        }$ & $  5.98 _{- 0.78 }^{+ 0.76 } $ & edge        \Tstrut\Bstrut\\ 
39          & $47          _{-7           }^{+8           }$ & $0.03        _{-0.16        }^{+0.17        }$ & $1.47        _{-0.04        }^{+0.09        }$ & $-1                                          $ & $ < 5.5                                      $ & $1.83        _{-0.38        }^{+0.36        }$ & $  3.18 _{- 0.72 }^{+ 0.69 } $ & -           \Tstrut\Bstrut\\ 
40          & $55          _{-8           }^{+9           }$ & $0.21        _{-0.15        }^{+0.15        }$ & $1.72        _{-0.26        }^{+0.76        }$ & $-1                                          $ & $4.8         _{-2.5         }^{+4.1         }$ & $2.35        _{-0.40        }^{+0.30        }$ & $  5.67 _{- 1.14 }^{+ 0.78 } $ & -           \Tstrut\Bstrut\\ 
41          & $118         _{-11          }^{+12          }$ & $0.02        _{-0.11        }^{+0.11        }$ & $2.28        _{-0.32        }^{+0.23        }$ & $2.58        _{-0.49        }^{+0.19        }$ & $16.8        _{-3.9         }^{+5.1         }$ & $3.53        _{-0.45        }^{+0.39        }$ & $  20.08 _{- 2.99 }^{+ 3.02 } $ & edge        \Tstrut\Bstrut\\ 
46          & $121         _{-11          }^{+13          }$ & $0.06        _{-0.10        }^{+0.10        }$ & $2.10        _{-0.09        }^{+0.20        }$ & $1.85        _{-0.02        }^{+0.03        }$ & $4.4         _{-2.0         }^{+2.3         }$ & $5.15        _{-0.64        }^{+0.59        }$ & $  20.88 _{- 2.94 }^{+ 2.88 } $ & line        \Tstrut\Bstrut\\ 
47          & $72          _{-9           }^{+10          }$ & $0.55        _{-0.13        }^{+0.13        }$ & $1.50        _{-0.15        }^{+0.07        }$ & $1.38        _{-0.84        }^{+0.32        }$ & $16.8        _{-3.6         }^{+4.5         }$ & $4.40        _{-0.72        }^{+0.63        }$ & $  9.79 _{- 1.88 }^{+ 1.80 } $ & edge        \Tstrut\Bstrut\\ 
48          & $44          _{-7           }^{+9           }$ & $0.03        _{-0.16        }^{+0.16        }$ & $1.71        _{-0.07        }^{+0.06        }$ & $-1                                          $ & $ < 6.8                                      $ & $1.94        _{-0.32        }^{+0.31        }$ & $  4.85 _{- 1.05 }^{+ 1.1 } $ & -           \Tstrut\Bstrut\\ 
55          & $52          _{-7           }^{+9           }$ & $0.02        _{-0.14        }^{+0.15        }$ & $0.69        _{-0.04        }^{+0.02        }$ & $-1                                          $ & $3.9         _{-1.2         }^{+1.4         }$ & $5.86        _{-1.03        }^{+1.08        }$ & $  1.80 _{- 0.38 }^{+ 0.35 } $ & -           \Tstrut\Bstrut\\ 
56          & $100         _{-11          }^{+12          }$ & $-0.02       _{-0.11        }^{+0.12        }$ & $0.83        _{-0.02        }^{+0.03        }$ & $-1                                          $ & $1.7         _{-0.7         }^{+0.8         }$ & $3.82        _{-0.41        }^{+0.60        }$ & $  1.69 _{- 0.26 }^{+ 0.25 } $ & -           \Tstrut\Bstrut\\ 
58          & $28          _{-6           }^{+7           }$ & $0.43        _{-0.15        }^{+0.20        }$ & $2.65        _{-0.06        }^{+0.10        }$ & $-1                                          $ & $17.6        _{-6.1         }^{+14.5        }$ & $8.34        _{-2.80        }^{+1.93        }$ & $  61.96 _{- 17.4 }^{+ 21.57 } $ & -           \Tstrut\Bstrut\\ 
59          & $295         _{-18          }^{+19          }$ & $0.61        _{-0.06        }^{+0.08        }$ & $1.00        _{-0.03        }^{+0.03        }$ & $0.79        _{-0.38        }^{+0.21        }$ & $16.8        _{-1.6         }^{+1.8         }$ & $16.63       _{-1.22        }^{+1.02        }$ & $  15.55 _{- 1.37 }^{+ 1.36 } $ & edge        \Tstrut\Bstrut\\ 
62          & $55          _{-8           }^{+9           }$ & $0.15        _{-0.17        }^{+0.19        }$ & $-1                                          $ & $2.76        _{-0.24        }^{+0.11        }$ & $47.0        _{-16.2        }^{+-13.4       }$ & $4.40        _{-0.92        }^{+0.72        }$ & $  50.16 _{- 12.13 }^{+ 11.98 } $ & edge        \Tstrut\Bstrut\\ 
67          & $205         _{-15          }^{+16          }$ & $0.16        _{-0.08        }^{+0.08        }$ & $0.35        _{-0.10        }^{+0.06        }$ & $0.38        _{-0.01        }^{+0.01        }$ & $1.3         _{-0.3         }^{+0.3         }$ & $6.54        _{-0.62        }^{+0.62        }$ & $  0.44 _{- 0.05 }^{+ 0.05 } $ & line        \Tstrut\Bstrut\\ 
69          & $223         _{-16          }^{+17          }$ & $0.15        _{-0.09        }^{+0.09        }$ & $-1                                          $ & $1.50        _{-0.91        }^{+0.90        }$ & $10.5        _{-1.8         }^{+2.0         }$ & $8.29        _{-0.69        }^{+0.65        }$ & $  17.00 _{- 1.82 }^{+ 1.77 } $ & edge        \Tstrut\Bstrut\\ 
70          & $215         _{-15          }^{+16          }$ & $0.28        _{-0.07        }^{+0.07        }$ & $0.72        _{-0.03        }^{+0.04        }$ & $0.76        _{-0.07        }^{+0.10        }$ & $5.1         _{-0.7         }^{+0.7         }$ & $10.80       _{-0.76        }^{+0.87        }$ & $  3.74 _{- 0.36 }^{+ 0.35 } $ & edge        \Tstrut\Bstrut\\ 
75          & $109         _{-12          }^{+13          }$ & $0.60        _{-0.10        }^{+0.14        }$ & $0.80        _{-0.01        }^{+0.04        }$ & $1.44        _{-0.65        }^{+0.28        }$ & $12.1        _{-2.0         }^{+2.5         }$ & $10.70       _{-1.64        }^{+1.24        }$ & $  5.16 _{- 0.80 }^{+ 0.75 } $ & edge        \Tstrut\Bstrut\\ 
82          & $123         _{-12          }^{+13          }$ & $0.24        _{-0.11        }^{+0.13        }$ & $4.06        _{-0.45        }^{+0.36        }$ & $-1                                          $ & $53.4        _{-11.1        }^{+12.1        }$ & $5.29        _{-0.76        }^{+0.76        }$ & $  154.38 _{- 23.74 }^{+ 22.2 } $ & -           \Tstrut\Bstrut\\ 
87          & $39          _{-7           }^{+8           }$ & $0.51        _{-0.13        }^{+0.18        }$ & $1.56        _{-0.05        }^{+0.10        }$ & $1.54        _{-0.07        }^{+0.13        }$ & $14.2        _{-3.5         }^{+4.8         }$ & $19.23       _{-3.66        }^{+3.63        }$ & $  14.03 _{- 3.64 }^{+ 3.61 } $ & edge        \Tstrut\Bstrut\\ 
89          & $39          _{-7           }^{+8           }$ & $0.13        _{-0.20        }^{+0.22        }$ & $1.01        _{-0.03        }^{+0.05        }$ & $1.52        _{-0.92        }^{+0.64        }$ & $6.9         _{-2.4         }^{+3.0         }$ & $4.28        _{-0.99        }^{+1.06        }$ & $  3.27 _{- 0.89 }^{+ 0.79 } $ & edge        \Tstrut\Bstrut\\ 
91          & $128         _{-12          }^{+13          }$ & $-0.10       _{-0.17        }^{+0.16        }$ & $1.06        _{-0.04        }^{+0.22        }$ & $0.92        _{-0.02        }^{+0.03        }$ & $1.5         _{-0.8         }^{+0.9         }$ & $18.78       _{-2.30        }^{+2.18        }$ & $  10.41 _{- 1.47 }^{+ 1.37 } $ & line        \Tstrut\Bstrut\\ 
94          & $67          _{-9           }^{+10          }$ & $0.52        _{-0.11        }^{+0.12        }$ & $4.08        _{-0.69        }^{+0.10        }$ & $2.72        _{-0.14        }^{+0.23        }$ & $105.4       _{-20.5        }^{+23.6        }$ & $6.50        _{-0.97        }^{+1.08        }$ & $  155.99 _{- 28.57 }^{+ 26.71 } $ & edge        \Tstrut\Bstrut\\ 
106         & $107         _{-11          }^{+12          }$ & $0.16        _{-0.12        }^{+0.13        }$ & $0.53        _{-0.03        }^{+0.02        }$ & $0.37        _{-0.02        }^{+0.02        }$ & $1.7         _{-0.4         }^{+0.5         }$ & $5.09        _{-0.63        }^{+0.52        }$ & $  0.33 _{- 0.05 }^{+ 0.05 } $ & line        \Tstrut\Bstrut\\ 
110         & $120         _{-16          }^{+17          }$ & $0.22        _{-0.14        }^{+0.17        }$ & $0.98        _{-0.08        }^{+0.04        }$ & $1.56        _{-0.94        }^{+0.33        }$ & $6.4         _{-1.5         }^{+1.8         }$ & $6.01        _{-0.91        }^{+0.93        }$ & $  5.24 _{- 0.78 }^{+ 0.71 } $ & edge        \Tstrut\Bstrut\\ 
115         & $40          _{-7           }^{+8           }$ & $0.52        _{-0.11        }^{+0.14        }$ & $0.46        _{-0.03        }^{+0.02        }$ & $0.66        _{-0.28        }^{+0.56        }$ & $5.2         _{-1.4         }^{+1.9         }$ & $5.76        _{-1.31        }^{+1.00        }$ & $  0.73 _{- 0.18 }^{+ 0.17 } $ & edge        \Tstrut\Bstrut\\ 
117         & $33          _{-6           }^{+7           }$ & $0.04        _{-0.19        }^{+0.20        }$ & $3.10        _{-0.60        }^{+0.80        }$ & $2.80        _{-1.68        }^{+0.35        }$ & $36.0        _{-13.4        }^{+16.3        }$ & $4.50        _{-1.00        }^{+1.36        }$ & $  54.26 _{- 15.84 }^{+ 14.63 } $ & edge        \Tstrut\Bstrut\\ 
119         & $46          _{-7           }^{+9           }$ & $0.30        _{-0.15        }^{+0.17        }$ & $0.54        _{-0.02        }^{+0.02        }$ & $-1                                          $ & $4.3         _{-1.2         }^{+1.5         }$ & $4.31        _{-0.96        }^{+0.76        }$ & $  0.76 _{- 0.17 }^{+ 0.16 } $ & -           \Tstrut\Bstrut\\ 
120         & $19          _{-5           }^{+6           }$ & $-0.09       _{-0.21        }^{+0.21        }$ & $0.95        _{-0.07        }^{+0.09        }$ & $-1                                          $ & $ < 2.1                                      $ & $4.12        _{-1.59        }^{+1.19        }$ & $  2.25 _{- 0.52 }^{+ 0.53 } $ & -           \Tstrut\Bstrut\\ 
122         & $82          _{-10          }^{+11          }$ & $0.29        _{-0.12        }^{+0.13        }$ & $1.29        _{-0.04        }^{+0.02        }$ & $0.98        _{-0.60        }^{+0.53        }$ & $9.6         _{-2.6         }^{+2.9         }$ & $6.28        _{-0.85        }^{+0.98        }$ & $  9.03 _{- 1.60 }^{+ 1.51 } $ & edge        \Tstrut\Bstrut\\ 
127         & $35          _{-6           }^{+7           }$ & $-0.02       _{-0.15        }^{+0.16        }$ & $0.05        _{-0.02        }^{+0.03        }$ & $-1                                          $ & $ < 1.1                                      $ & $2.23        _{-0.51        }^{+0.42        }$ & $ (1.6_{-0.3}^{+0.4})$e-3     & -        \Tstrut\Bstrut\\ 
130         & $26          _{-6           }^{+7           }$ & $-0.07       _{-0.32        }^{+0.33        }$ & $-1                                          $ & $-1                                          $ & $-1                                          $ & $6.28        _{-1.80        }^{+1.48        }$ & $ -1         $ & -           \Tstrut\Bstrut\\ 
131         & $39          _{-7           }^{+8           }$ & $0.05        _{-0.21        }^{+0.23        }$ & $1.68        _{-0.15        }^{+0.12        }$ & $1.53        _{-0.47        }^{+0.71        }$ & $11.8        _{-4.6         }^{+6.2         }$ & $4.09        _{-0.94        }^{+0.91        }$ & $  11.03 _{- 3.00 }^{+ 2.81 } $ & edge        \Tstrut\Bstrut\\ 
135         & $34          _{-6           }^{+8           }$ & $0.05        _{-0.18        }^{+0.20        }$ & $-1                                          $ & $-1                                          $ & $-1                                          $ & $3.72        _{-0.76        }^{+0.83        }$ & $ -1         $ & -           \Tstrut\Bstrut\\ 
137         & $50          _{-8           }^{+9           }$ & $-0.04       _{-0.18        }^{+0.19        }$ & $1.48        _{-0.76        }^{+1.10        }$ & $-1                                          $ & $ < 18.7                                     $ & $4.70        _{-0.87        }^{+0.86        }$ & $  8.96 _{- 3.31 }^{+ 2.58 } $ & -           \Tstrut\Bstrut\\ 
140         & $40          _{-7           }^{+9           }$ & $0.02        _{-0.23        }^{+0.24        }$ & $0.92        _{-0.04        }^{+0.05        }$ & $0.82        _{-0.44        }^{+0.74        }$ & $ < 4.0                                      $ & $2.14        _{-0.45        }^{+0.48        }$ & $  1.06 _{- 0.29 }^{+ 0.29 } $ & edge        \Tstrut\Bstrut\\ 
192         & $138         _{-15          }^{+17          }$ & $1.0                                         $ & $1.76        _{-0.04        }^{+0.06        }$ & $2.20        _{-0.28        }^{+0.15        }$ & $77.4        _{-12.5        }^{+15.5        }$ & $10.93       _{-1.41        }^{+1.14        }$ & $  60.94 _{- 14.11 }^{+ 13.66 } $ & edge        \Tstrut\Bstrut\\ 
196         & $76          _{-9           }^{+10          }$ & $1.0                                         $ & $0.54        _{-0.01        }^{+0.02        }$ & $0.41        _{-0.25        }^{+0.05        }$ & $28.9        _{-5.8         }^{+10.2        }$ & $4.67        _{-0.67        }^{+0.43        }$ & $  1.67 _{- 0.37 }^{+ 0.48 } $ & edge        \Tstrut\Bstrut\\ 
200         & $162         _{-15          }^{+16          }$ & $1.0                                         $ & $1.42        _{-0.04        }^{+0.03        }$ & $1.59        _{-0.54        }^{+0.59        }$ & $3.4         _{-1.2         }^{+1.4         }$ & $6.40        _{-0.75        }^{+0.58        }$ & $  10.48 _{- 1.29 }^{+ 1.23 } $ & edge        \Tstrut\Bstrut\\ 
201         & $55          _{-8           }^{+9           }$ & $1.0                                         $ & $0.86        _{-0.05        }^{+0.02        }$ & $1.15        _{-0.21        }^{+0.08        }$ & $34.1        _{-6.9         }^{+10.5        }$ & $6.41        _{-1.15        }^{+0.86        }$ & $  6.12 _{- 1.48 }^{+ 1.63 } $ & edge        \Tstrut\Bstrut\\ 
207         & $78          _{-10          }^{+11          }$ & $1.0                                         $ & $0.14        _{-0.04        }^{+0.03        }$ & $0.14        _{-0.09        }^{+0.58        }$ & $6.3         _{-1.2         }^{+1.9         }$ & $9.50        _{-1.50        }^{+1.43        }$ & $  0.10 _{- 0.02 }^{+ 0.02 } $ & edge        \Tstrut\Bstrut\\ 
220         & $34          _{-7           }^{+8           }$ & $1.0                                         $ & $1.57        _{-0.08        }^{+0.06        }$ & $2.23        _{-0.34        }^{+0.21        }$ & $60.5        _{-17.3        }^{+27.9        }$ & $6.95        _{-1.41        }^{+1.54        }$ & $  26.87 _{- 10.3 }^{+ 10.58 } $ & edge        \Tstrut\Bstrut\\ 
221         & $30          _{-6           }^{+7           }$ & $1.0                                         $ & $-1                                          $ & $1.37        _{-0.28        }^{+0.08        }$ & $58.8        _{-18.6        }^{+25.1        }$ & $8.21        _{-1.95        }^{+1.58        }$ & $  25.15 _{- 9.70 }^{+ 9.71 } $ & edge        \Tstrut\Bstrut\Tstrut\Bstrut\\ 
\enddata

\tablecomments{(1): X-ray ID; (2): net counts in the full (0.5-7 keV) band from the extracted spectra, where the errors were computed according to \cite{gehrels86} and correspond to the 1$\sigma$ level in Gaussian statistics; (3) HR calculated by \citetalias{nanni20}, where we put 1.0 for sources where only hard counts were detected; (4) photometric redshifts obtained from the SED fitting procedure; (5) X-ray redshift solutions; (6): column density in units of $10^{22}$ cm$^{-2}$; (7): observed 0.5-7 keV flux in units of $10^{-15}$ erg s$^{-1}$ cm$^{-2}$; (8): intrinsic, absorption-corrected luminosity in the 2-10 keV rest-frame band in units of $10^{43}$ erg s$^{-2}$ and (9): X-ray feature on which the $z_X$ is based, 'line' for the 6.4 keV Fe K$\alpha$ line and 'edge' for the 7.1 keV Fe absorption edge and associated photoelectric cut-off. '-1' indicates a non derived quantity. Uncertainties are reported at 1$\sigma$ confidence level.}
\label{my_results}
\end{deluxetable*}

\begin{appendix}
\restartappendixnumbering
\section{HR dependencies}\label{app_A}
In recent years, the \textit{Chandra} ACIS camera has experienced a decline in the effective area. Some gaseous material has settled on the cold ACIS optical blocking filters, reducing the photon collection efficiency of the instrument especially at soft X-ray energies (see the Chandra Proposers' Observatory Guide, December 2019). This has a large impact in the analysis of quantities like the $HR$, which takes into account both soft and hard count rates of the observed sources.
In this paragraph we analyze the ACIS-I ARF degradation and how this affects the hardness ratio analysis. In this regard, we used the aimpoint ARFs available on the dedicated CXC web-page\footnote{\url{http://cxc.harvard.edu/caldb/prop\_plan/imaging/}}.

\begin{figure}[t!]
    \begin{center}
        \includegraphics[width=0.47\textwidth]{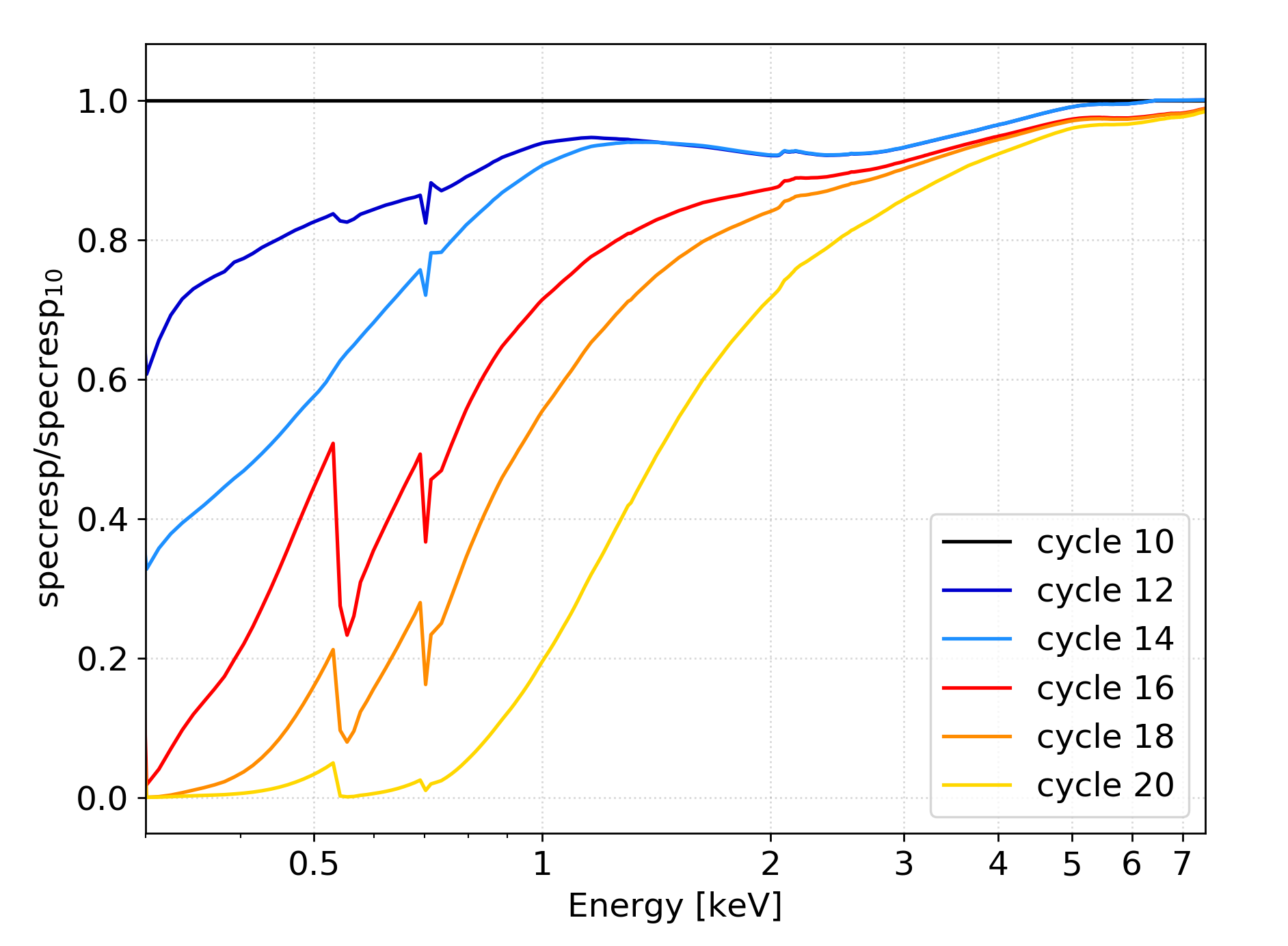}\hfil
        \caption{ACIS-I effective area degradation through years. We show the ARF ratios between cycles 12 (blue), 14 (light blue), 16 (red), 18 (orange), 20 (yellow) and the cycle 10 (black).}
        \label{arf_ratio}
    \end {center}
\end{figure}

In Figure \ref{arf_ratio} we show the ratio between the ARFs of different \textit{Chandra} cycles and that of cycle 10. Below 2 keV the response is dramatically decreasing with increasing cycles (i.e., years), while at higher energies the degradation is less significant. As an example, the cycle 20 response at 1 keV is decreased by 80\% and 65\% compared to cycles 10 and 18, respectively, while at 5 keV it is decreased only by 4\% and $<1\%$ compared to the same cycles.
In particular, this trend does not allow us to compare our results with literature (e.g., \citealt{tozzi01,szokoly04}), when the higher photon collection efficiency below 2 keV produced lower $HR$ values for a given spectral shape.

\begin{figure}[t!]
	\begin{center}
		\includegraphics[width=0.47\textwidth]{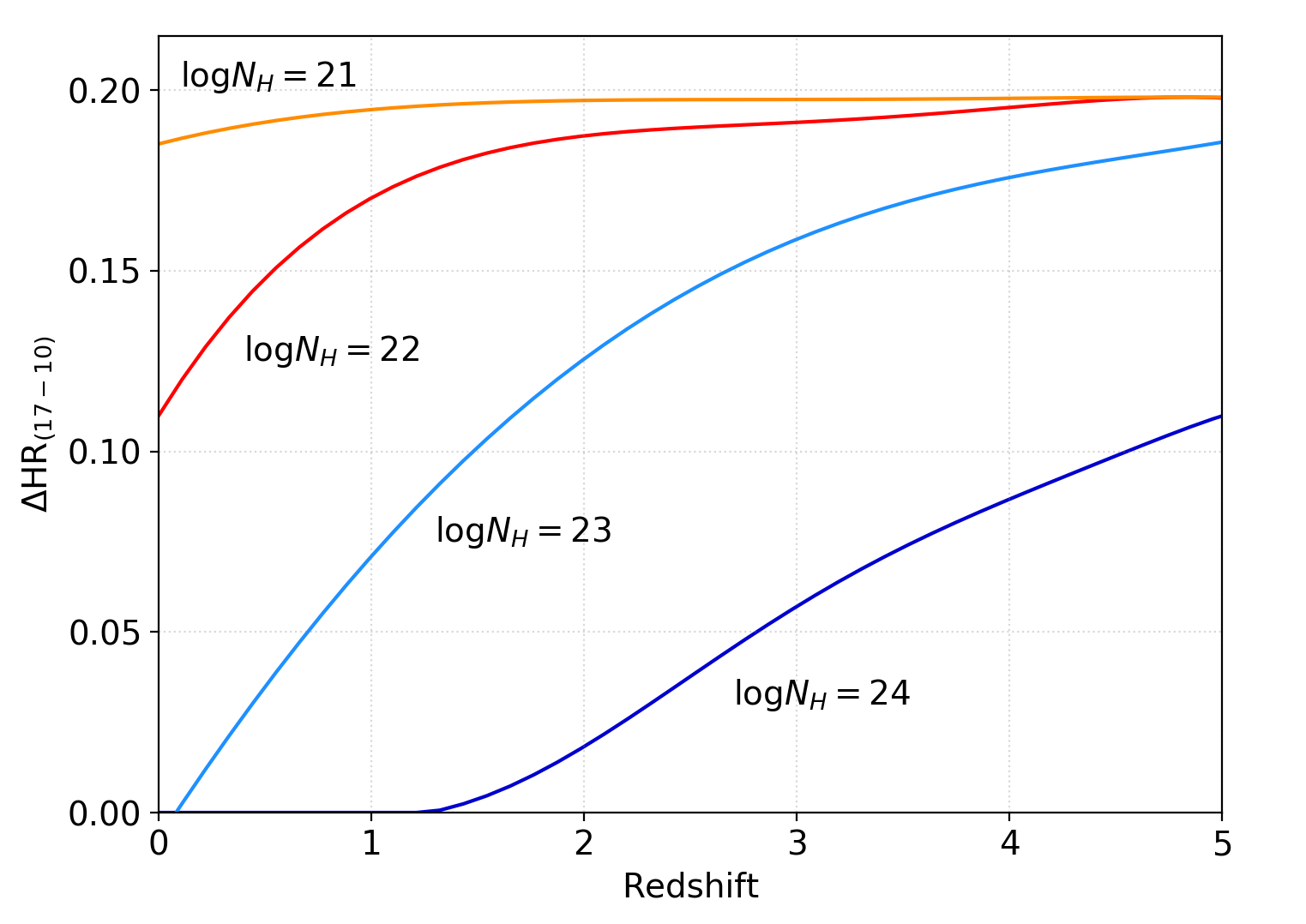}\hfil
		\caption{Hardness ratios difference between cycle 17 (this work) and cycle 10, as a function of redshift and for different absorption column densities in color code. We assumed a simple absorbed power-law model with a fixed $\Gamma=1.9$.}
		\label{arf_diff}
	\end {center}
\end{figure}

In Figure \ref{arf_diff} we compare the differences of $HR$ obtained using the response matrices of cycle 17 (this work) and cycle 10. The difference is lower for the most obscured AGN (log$N_H=23,24$), especially at low redshift. This effect is due to the fact that objects with high $N_H$ have almost only hard X-ray emission, which is collected by the part of the detector less affected from the contaminating material. On the contrary, the difference is larger for sources with low $N_H$ (log$N_H=21,22$) because they also have soft X-ray emission, which is more affected by the detector contamination. Increasing the redshift, part of the rest-frame hard emission is redshifted to lower energies, increasing the differences for sources with high $N_H$, while this effect is mitigated for sources with low $N_H$, because of the flatter spectral shape.

\section{Tests for different spectral models} \label{app_B}
\subsection{HR trends}\label{app_B1}
\restartappendixnumbering

\begin{figure}[th!]
    \begin{center}
        \includegraphics[width=0.5\textwidth]{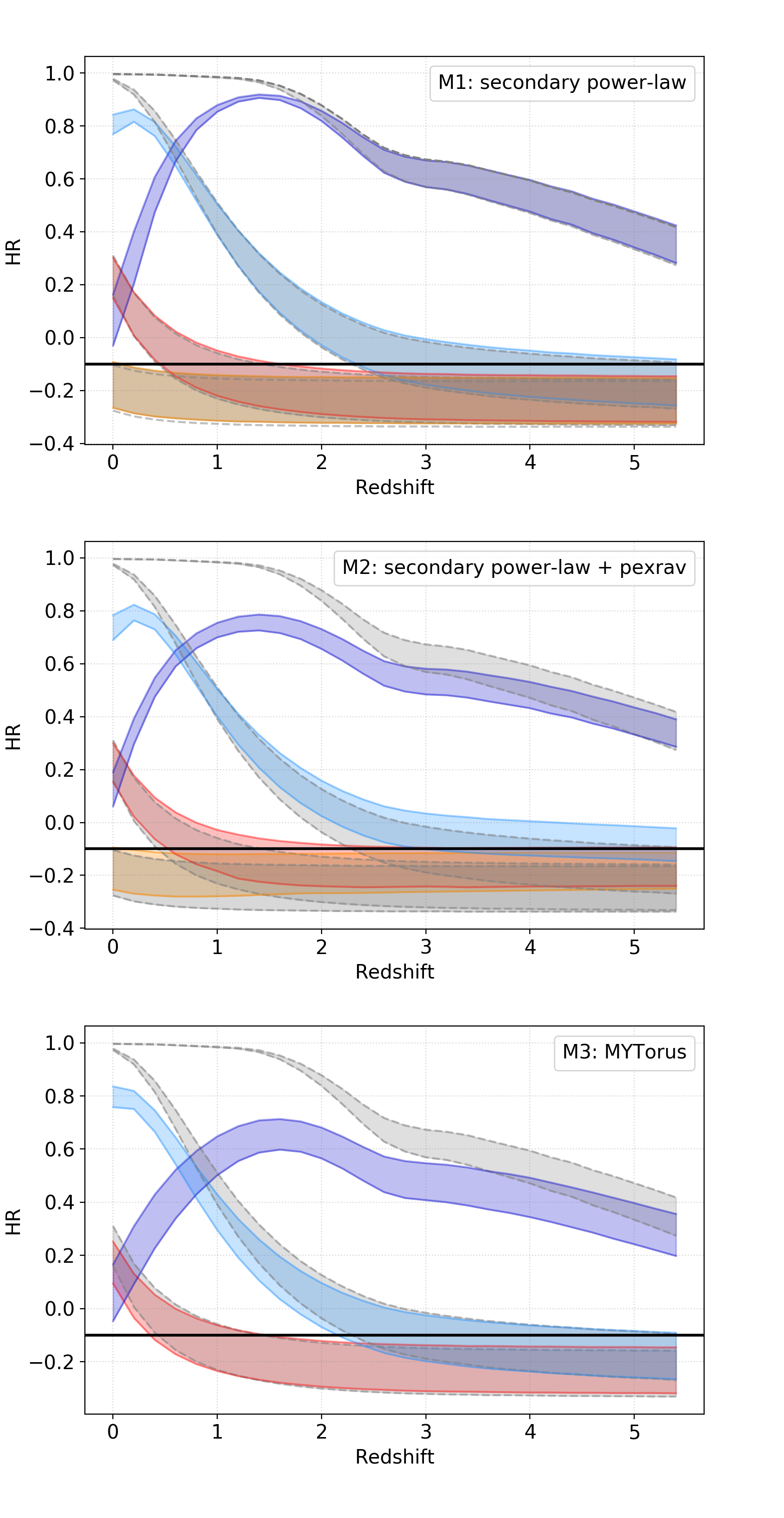}\hfil
        \caption{$HR$ trends as a function of redshift and for different absorption column densities (log$N_H$=21,22,23, and 24 in orange, red, light blue, and blue, respectively). For each $N_H$ the shaded area represents $HR$ values obtained for different $\Gamma$, in the range 1.7$-$2.1. From top to bottom, trends derived with different models are shown: main absorbed power-law plus a redshifted secondary power-law (M1; top panel); M1 plus a reflection component (M2; middle panel); MYTorus model with reflection and secondary power-law (M3; bottom panel). For comparison, the grey shaded areas refer to the single power-law model (M0) used in this paper (see Figure \ref{tozzi_HR}). The black solid lines represent the chosen $HR>-0.1$ threshold.}
        \label{hr_tests}
    \end {center}
\end{figure}

To validate our $HR>-0.1$ selection criterion for obscured AGN, we tested how adopting different spectral shapes affects the $HR$-redshift curves for different $N_H$ values. We remind that the model used in this paper is a simple absorbed power-law (hereafter M0; Figure \ref{tozzi_HR}). The simulations were performed as discussed in Section \ref{sample_selection}. In each model, the mean Galactic absorption at the J1030 field position ($N_H=2.6\times 10^{20}$ cm$^{-2}$) was considered.

We modelled the soft excess emission adding a secondary redshifted power-law (\texttt{zpowerlw}) to M0. The redshift parameter was linked between the components and no intrinsic absorption was applied to the secondary power-law, assuming it is scattered emission into the line of sight. The photon index $\Gamma$ of the two power-laws was fixed to 1.9. The soft excess contribution to the main continuum is observed to be $<$10\%, with typical values of 1-3 percent (e.g., \citealt{gilli07, ricci17}), and it was considered by adding a multiplicative 3\% constant to the secondary power-law.
The results for this model (hereafter, M1) are shown in Figure \ref{hr_tests} (top panel). Compared to the M0 curves, there is a strong $HR$ decrease for log$N_H$=24 at $z<2$ (up to $\Delta$\textit{HR}$\sim$0.5-1 at $z<0.5$), a small decrease for log$N_H$=23 at $z<0.5$ ($\Delta$\textit{HR}$\sim$0.1-0.2), and no significant effects for log$N_H$=22-21. This is due to the fact that, as the $N_H$ increases, the soft rest-frame main emission is more and more depressed by the absorption, and therefore the soft excess becomes dominant at these energies. This effect is diluted at high redshift, when the primary hard rest-frame emission is redshifted enough to cover the observed-frame soft band.

The reflection is modelled by adding a \texttt{pexrav} component (\citealt{magdziarz95}) to M1. The $R$ value was fixed to -1 (e.g., \citealt{marchesi16b}) to simulate pure reflection, while photon index and normalization were linked to the main power-law component. Inclination angle and high-energy cut-off were set to the default values $\theta=60\degree$ and $E=100$ keV, respectively. The results for this model (hereafter, M2) are shown in Figure \ref{hr_tests} (middle panel).
There are no major changes in the $HR$ values as a function of redshift, compared to the double power-law model. For log$N_H$=21, 22 and 23 the $HR$ slightly increases at $z\gtrsim1$ ($\Delta$\textit{HR}$<$0.1), while for log$N_H$=24 there is an increase at $z<0.5$ and a decrease at higher redshift ($\Delta$\textit{HR}$\lesssim$0.1). The \texttt{pexrav} model adds flux both in the soft and in the hard rest-frame bands. Therefore, as a function of redshift and $N_H$, there is a combination of reflection and soft excess which subtly changes the $HR$ values.

For comparison with M2, in Figure \ref{hr_tests} (bottom panel) we show the results obtained with the MYTorus model (\citealt{murphy09}). It adopts an azimuthally-symmetric toroidal shape for the obscuring material, with a fixed half opening angle of 60$\degree$. The torus material is assumed to be neutral, cold and uniform. The reflection is included within this model, but with a more physically motivated treatment, while the soft excess is modelled with a secondary redshifted power-law. We refer to this model as M3. We fixed $\Gamma=1.9$ for all the components and the inclination angle between the observer and the torus axis to $\theta=75\degree$. In MYTorus, log$N_H$ varies in the range 22-25, so the log$N_H$=21 curve is not reported.
The curve for log$N_H$=22 is very similar to those of M0 and M1. The log$N_H$=23 trend is similar to that of M1, and for log$N_H$=24 the $HR$ values are lower than those of all the other methods (e.g., $\Delta$\textit{HR}$\lesssim$0.1 compared to M2), with a larger spread for the chosen $\Gamma$ values. 
Interestingly, there is not the same $HR$ increase at $z\gtrsim1$ for log$N_H$=22 and 23 that we observe for M2, due to the different reflection treatment; see also \citet{marchesi20} for a comparison between the different shapes of \texttt{pexmon} (\citealt{nandra07}), that includes the \texttt{pexrav} model, and BORUS (\citealt{balankovic18}), a physically self consistent torus model.

In summary, as shown in Figure \ref{hr_tests}, the $HR=-0.1$ threshold (black solid line in each panel) remains valid. In fact, with an $HR>-0.1$, we avoid unobscured sources at any redshift.

\subsection{Redshift match \%}\label{app_B2}

We intend to demonstrate here that a simple absorbed power-law (M0) is a reasonable model to obtain reliable X-ray redshifts estimates. To prove it, we performed several run of simulations similar to those described in Section \ref{semitheor_sim}, using the M3 model described in Appendix \ref{app_B1}. The explored parameter space is log$N_H$=22, 23, 24, and $z$=0.5, 1, 2, 3, 4. We varied the power-law normalization to obtain sources with net counts in the ranges 10-100 (low regime) and 100-1000 (high regime). For each parameter combination, 100 spectra were simulated.
The simulated spectra were then fitted with M0 and the three models presented in Appendix \ref{app_B1}: M1 accounts for the soft excess, M2 and M3 also include the reflection. Compared to M0, treated as in Section \ref{semitheor_sim}, the additional free parameters in M1-M3 were the secondary power-law normalization, constrained to be $<10\%$ of the primary power-law, and the scattering normalization in M3.

\begin{figure}[!t]
    \begin{center}
        \includegraphics[width=0.5\textwidth]{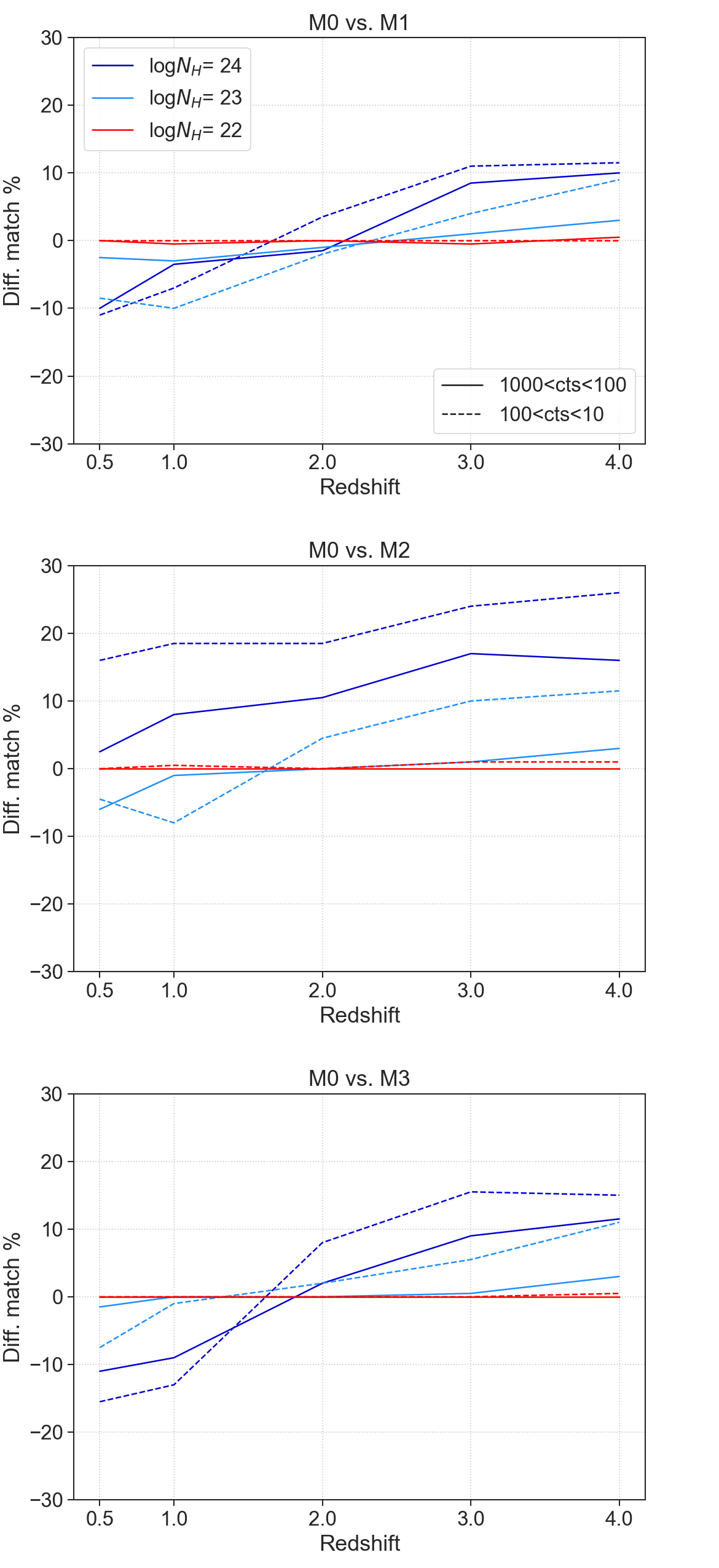}\hfil
        \caption{Match percentage difference between M0 and the chosen complex models: M1 (top panel), M2 (middle panel), and M3 (bottom panel), as a function of redshift. Sources with log$N_H$=22, 23, and 24 are showed in red, light blue and blue, respectively. The dashed and solid lines represent sources with net counts ranges 10-100 and 100-1000, respectively.}
        \label{matchperc_tests}
    \end {center}
\end{figure}
In Figure \ref{matchperc_tests} we show the difference of the match percentage (Equation \ref{matcheq}) between the results obtained with M0 and the three complex models. Positive values indicate that M0 is more effective in recovering the simulated source redshift, while negative values indicate that complex models are more efficient.
No clear trends are found for log$N_H$=22. For log$N_H$=23 and 24 there are clear trends as a function of redshift. When compared to M1-M3, M0 is penalized below $z\sim$1-2, while at higher redshift it is more effective. This is because the contribution of the secondary power-law, present in all the complex models, influences the fit for sources at low redshift, especially where the primary continuum is strongly extinguished (log$N_H>22$) in the soft band. At higher redshift, instead, this contribution is diluted by the redshifted main power-law. 
The reason for these discrepancies is attributed to the increased number of parameters in M1-M3. In particular, the secondary power-law normalization plays an important role, even if constrained to observed values. In fact, the spectral fit may interpret a noise fluctuation as a real spectral shape, fitting a wrong normalization and then a wrong redshift. The more complex is the model spectral shape, the higher is the risk of misinterpretation.
These differences are more evident for higher $N_H$, as discussed in Appendix \ref{app_B1}, and for the low count regime, because of the poor spectral quality.

In general, it is clear that for log$N_H$=22 there are no significant differences between the models ($<2\%$), as well as for log$N_H$=23 in the high-count regime ($<5\%$). 
For log$N_H$=24, and log$N_H$=23 in the low-count regime, the differences may instead be non-negligible. For log$N_H$=23 in the low-count regime the differences are $<$10\%, while for log$N_H$=24 they are $<$10\% and $<$15\% in the high and low-count regimes, respectively.
Overall, M0 gives better results above $z\sim$1-2 for all models, and is preferable against M2 for log$N_H$=24 at any redshift.
Below $z\sim$1-2, M0 is penalized for the remaining cases, with differences within $\sim$10\% except for M3 in the low counts regime, log$N_H$=24 and $z<1$, where there is a $\sim$15\% difference. However, for M3 there may be a bias introduced by the simulated model, which is the same as the fitted one, that may produce a larger difference in the match percentage against M0.

Given these results, we can safely say that using an absorbed power-law model is a reasonable assumption for the redshift estimate of obscured AGN. The main spectral features, such as the Fe 6.4 keV emission line, the Fe 7.1 keV absorption edge and the photoelectric absorption cut-off, are included in the absorbed power-law model and are those that drive the redshift solutions. Therefore, for a limited photon statistics, introducing more complex models not only does not substantially improve the results, but may also introduce degeneracies due to the possible misidentifications of complex features.

\end{appendix}

%\clearpage
\bibliography{sample63}{}
\bibliographystyle{yahapj}



\end{document}